\documentclass[11pt]{article} 
     \usepackage{amsmath}
  \usepackage{graphicx}
  \usepackage{color}
  \usepackage{subfigure}
  \usepackage{tikz}
  \newlength{\abstractwidth}
 \usepackage[numbers,sort&compress]{natbib}

 \usepackage{amsmath,amssymb}
\usepackage{graphicx}
\usepackage{caption}
\usepackage{lipsum}
\usepackage[utf8]{inputenc}

\usepackage{upgreek}

\usepackage[colorlinks = true,
            linkcolor = blue,
            urlcolor  = blue,
            citecolor = blue,
            anchorcolor = blue]{hyperref}

\hypersetup{colorlinks=true,linkcolor=blue}

\usepackage[title]{appendix}
\usepackage{comment}
\usepackage{braket}

\def\XXint#1#2#3{{\setbox0=\hbox{$#1{#2#3}{\int}$}
     \vcenter{\hbox{$#2#3$}}\kern-.5\wd0}}

\makeatletter
\def\blfootnote{\xdef\@thefnmark{}\@footnotetext}
\makeatother

  \newcommand{\be}{\begin{equation}}
  \newcommand{\bea}{\begin{eqnarray}}
  \newcommand{\eea}{\end{eqnarray}}
  \newcommand{\beq}{\begin{equation}}
  \newcommand{\ee}{\end{equation}}
  \newcommand{\eeq}{\end{equation}}

  \def\ba{\begin{eqnarray}}
  \def\ea{\end{eqnarray}}

 \def\simleq{\; \raise0.3ex\hbox{$<$\kern-0.75em
      \raise-1.1ex\hbox{$\sim$}}\; }
 \def\simgeq{\; \raise0.3ex\hbox{$>$\kern-0.75em
      \raise-1.1ex\hbox{$\sim$}}\; }

\begin{document}

\begin{titlepage}
  \bigskip

  \bigskip\bigskip

  \bigskip

\begin{center}
{\Large \bf{}}
 \bigskip
{\Large \bf {Spectral form factor for free large $N$ gauge theory and strings}} 
\bigskip
\bigskip
   \bigskip
\bigskip
\end{center}

  \begin{center}

 \bf {Yiming Chen}
  \bigskip \rm
\bigskip
 
 \rm

Jadwin Hall, Princeton University,  Princeton, NJ 08540, USA\\

  \end{center}

 \bigskip\bigskip

\begin{abstract}

We investigate the spectral form factor in two different systems, free large $N$ gauge theories and highly excited string gas.  In both cases, after a rapid decay of the spectral form factor at early time, new contributions come in, preventing the spectral form factor from ever becoming exponentially small. We consider $U(N)$ gauge theories with only adjoint matter and compute the spectral form factor using a matrix integral of the thermal holonomy $U$. The new saddles differ from the early time saddle by preserving certain subgroups of the center symmetry. For a gas of strings, the short time decay of the spectral form factor is governed by the continuous Hagedorn density of states, which can be associated to the  thermal winding mode with winding number $\pm1$. We show that the rise of the spectral form factor comes from other winding modes that also carry momentum along the time direction. We speculate on the existence of a family of classical solutions for these string modes, similar to the Horowitz-Polchinski solution.

We review a similar problem for black holes. In particular, we examine the Kontsevich-Segal criterion on complex black holes that contribute to the spectral form factor. In the canonical ensemble quantity $Z(\beta+it)$, the black hole becomes unallowed at $t\sim \mathcal{O}(\beta)$. A way to avoid this is to consider the microcanonical ensemble, where the black hole stays allowable.

\end{abstract}
\bigskip \bigskip \bigskip
\blfootnote{ymchen.phys@gmail.com}

  \end{titlepage}

   \tableofcontents


\section{Introduction and overview}

In recent years, progress has been made on understanding how semiclassical gravity encodes information about the microscopic degrees of freedom \cite{Shenker:2013pqa,Penington:2019npb,Almheiri:2019psf,Almheiri:2019hni,Almheiri:2019qdq,Penington:2019kki,Saad:2019lba}. A quantity which played an important role in recent discussions is the spectral form factor \cite{2010qsc,Papadodimas:2015xma}, defined as the square of the analytically continued partition function
\begin{equation}\label{Zdef}
	Z(\beta+ it) = \sum_{n} e^{-(\beta+it)E_n} .
\end{equation}
It is a useful quantity since as we increase $t$, it probes finer and finer detail of the microscopic spectrum, while on the other hand, the partition function is a natural quantity to compute from semiclassical gravity \cite{PhysRevD.15.2752}.

Although we will be mostly studying the spectral form factor in systems \emph{not} directly related to gravity, we find it useful to begin with a short review of the black hole problem, highlighting some lesser appreciated aspects and the kind of questions we will be asking.

For reasons we will explain, rather than considering the quantity $|Z(\beta+it)|^2$, we will instead focus on a variant of the spectral form factor $|Y_{E,\Delta} (t) |^2$ \cite{Stanfordunpub,Gharibyan:2018jrp}, where $Y_{E,\Delta}(t)$ is defined with a Gaussian window around energy $E$ with width $\Delta$,
\begin{equation}\label{Ydef}
	Y_{E,\Delta} (t) =  \sum_{n} e^{- \frac{(E_n - E)^2}{2\Delta^2}} e^{-iE_n t} .
\end{equation}
There are several benefits of considering $Y_{E,\Delta} (t)$ rather than $Z(\beta+it)$. First, if we consider $Z(\beta+it)$ for a higher dimensional AdS black hole, as we increase $t$, due to the cancellation among high energy states from the phases in (\ref{Zdef}), the temperature effectively lowers and at some $t \sim \mathcal{O}(\beta)$, the quantity will be dominated by the thermal AdS geometry rather than the black hole \cite{Cotler:2016fpe,Copetti:2020dil}. On the other hand, the thermal AdS will not contribute in the microcanonical quantity $Y_{E,\Delta} (t)$ if one sets the energy high enough, so one can use it to study the black hole at a longer time scale. 

Closely related, in section \ref{sec:bh}, we point out another reason why $Y_{E,\Delta} (t)$ is better based on the Kontsevich-Segal allowability criterion of complex metrics \cite{Kontsevich:2021dmb,Witten:2021nzp}. The upshot is that the complex black hole geometry that computes $Z(\beta+it)$ becomes unallowable at $t\sim \mathcal{O}(\beta)$. On the other hand, it stays allowable in $Y_{E,\Delta} (t)$.

The second main benefit of considering $Y_{E,\Delta} (t)$ over $Z(\beta+it)$, perhaps more essential for our discussion, is the property that $|Y_{E,\Delta} (t) |^2$ decays much faster compared with $|Z(\beta+it)|^2$ at $t\gtrsim \beta$, therefore is better in exposing other contributions to the spectral form factor \cite{Gharibyan:2018jrp}. To see the rapid decay of $|Y_{E,\Delta} (t) |^2$, consider a simple example where the density of states $\rho(E) \sim e^{S(E)}$ is almost flat within the energy window, and replace the discrete sum in (\ref{Ydef}) by a continuous integral. We get
\begin{equation}\label{decay}
	|Y_{E,\Delta} (t)|^2 \approx \left| \int d\tilde{E}\, e^{- \frac{(\tilde{E} - E)^2}{2\Delta^2}} e^{-i\tilde{E} t}  \rho(\tilde{E}) \right|^2 \sim e^{2S(E) - \Delta^2 t^2}.
\end{equation}
\begin{figure}[t!]
\begin{center}
\includegraphics[scale=0.27]{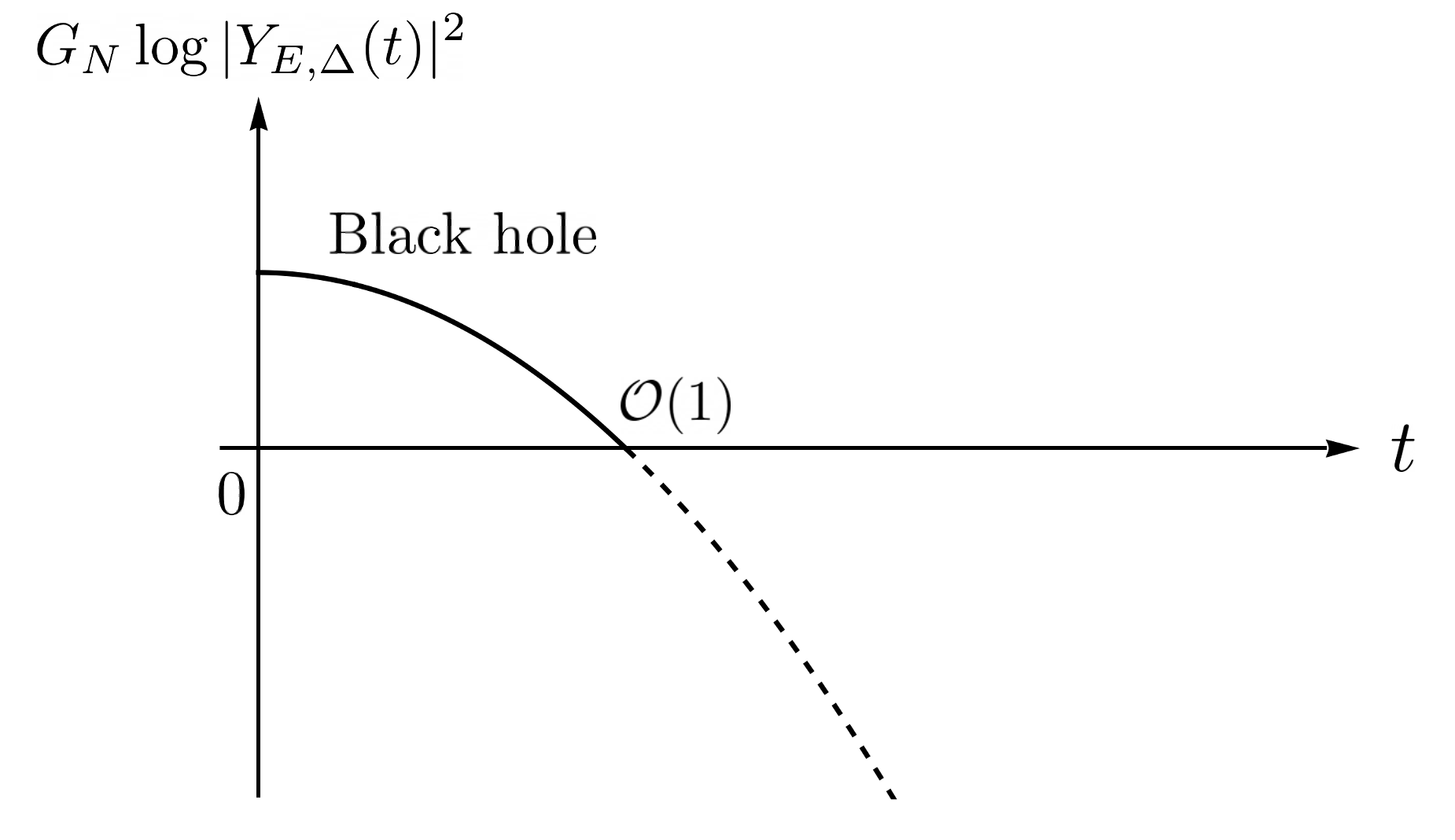}
\caption{The spectral form factor $|Y_{E,\Delta}(t)|^2$ for a black hole. The semiclassical answer becomes exponentially small at $t\sim \mathcal{O}(1)$. }
\label{fig:BH}
\end{center}
\end{figure}
For a non-flat but slowly varying density of states $\rho(E)$ one gets a similar decay. In particular, a semiclassical computation using the black hole geometry leads to a similar decay, which we will discuss with more detail in section \ref{sec:bh}. We sketch the answer in fig. \ref{fig:BH}. On contrary, at $t\gg \beta$, the decay of $|Z(\beta+it)|^2$ is determined by the spectral density near the edge of the spectrum. For example, for a random matrix that has a square root edge in the spectrum, $|Z(\beta+it)|^2$ decays as $1/t^3$ at $t\gg \beta$ \cite{Cotler:2016fpe}.

For an AdS black hole, it is convenient to choose $\Delta \sim \mathcal{O}(1/\sqrt{G_N})$ and (\ref{decay}) suggests that when $t \sim \mathcal{O}(1)$, $|Y_{E,\Delta}(t)|^2$ becomes exponentially small, of order $\mathcal{O}(e^{-1/G_N})$. We note that this would be very surprising from the point of view of (\ref{Ydef}), since each term in the discrete sum contributes an order one number. Getting an exponentially small final answer requires a delicate cancellation between an exponential number of states, though in principle possible, seems very unlikely for a chaotic system like a black hole.\footnote{We are not claiming that this is impossible. One can easily construct special spectra that achieve this.} We view this as an \emph{early time challenge to the discreteness of the spectrum}.

The puzzle is what geometry replaces the black hole. This is a well-known problem. Above we are merely pointing out a lesser appreciated aspect that it is a puzzle already at order one time, when looking at suitable quantities. It has been proposed that under some average, possibly over time, the double cone wormhole geometry \cite{Saad:2018bqo,Mahajan:2021maz,Cotler:2021cqa} gives an order one increasing contribution called the ``ramp". This is most well understood in lower dimensional gravity models like JT gravity \cite{JACKIW1985343,TEITELBOIM198341}, which has an ensemble interpretation \cite{Saad:2019lba}.\footnote{The inclusion of spacetime wormholes in the gravity path integral has spurred a lot of recent discussion (see for example \cite{Stanford:2020wkf,Cotler:2020ugk,Saad:2021uzi,Belin:2021ibv,Verlinde:2021jwu,Collier:2022emf,Johnson:2022wsr,Schlenker:2022dyo} and more in a recent review \cite{Bousso:2022ntt}), in particular due to the factorization puzzle \cite{Maldacena:2004rf}. We will not touch on this issue here.} In systems without an average, it was found in toy models that there can be certain one boundary contributions called ``half-wormhole" \cite{Saad:2021rcu,Mukhametzhanov:2021nea,Blommaert:2021fob,Garcia-Garcia:2021squ}, but the general mechanism is unclear at the moment.

In this paper we will discuss how analogous puzzles are resolved in much simpler quantum systems, including large $N$ free gauge theories and string gases at high energies. Of course, it would be quite boring if we just compute $|Y_{E,\Delta}(t)|^2$ directly from the spectrum. After all, these are systems that we in principle know the microstates and in the end we certainly will not find problems with discreteness. However, the story becomes interesting since in both cases, the computation can be phrased in terms of quantities that have some ``geometrical" flavor while the discreteness of the spectrum becomes no longer manifest. We will see that the ``geometries" that are commonly known in these systems \emph{do} lead to a rapid decay of $|Y_{E,\Delta}(t)|^2$ at short time, and we will study how new ``geometries" come in and prevent $|Y_{E,\Delta}(t)|^2$ from decaying to exponentially small.

For now, let's simply point out the meaning of ``geometries" in both systems and we shall discuss them in much more detail later. For the large $N$ gauge theories, the analogs of ``geometries" are the different eigenvalue distributions of the thermal holonomy. For a gas of string, the analogs of ``geometries" are different string winding modes on a thermal manifold, or more literally, the classical solutions of them such as the Horowitz-Polchinski solution \cite{Horowitz:1997jc}.

Of course, we should emphasize that these systems have very different spectra compared to a many-body chaotic system like the black hole. For both the free gauge theory and a non-interacting string gas, their spectra have large degeneracies and do not contain random matrix behavior. Therefore we expect the exact mechanisms that the black hole solves the puzzle to be quite different. Nonetheless, we will discuss some possible ways to improve our understanding towards  systems with many-body chaos.

\begin{figure}[t!]
\begin{center}
\includegraphics[scale=0.3]{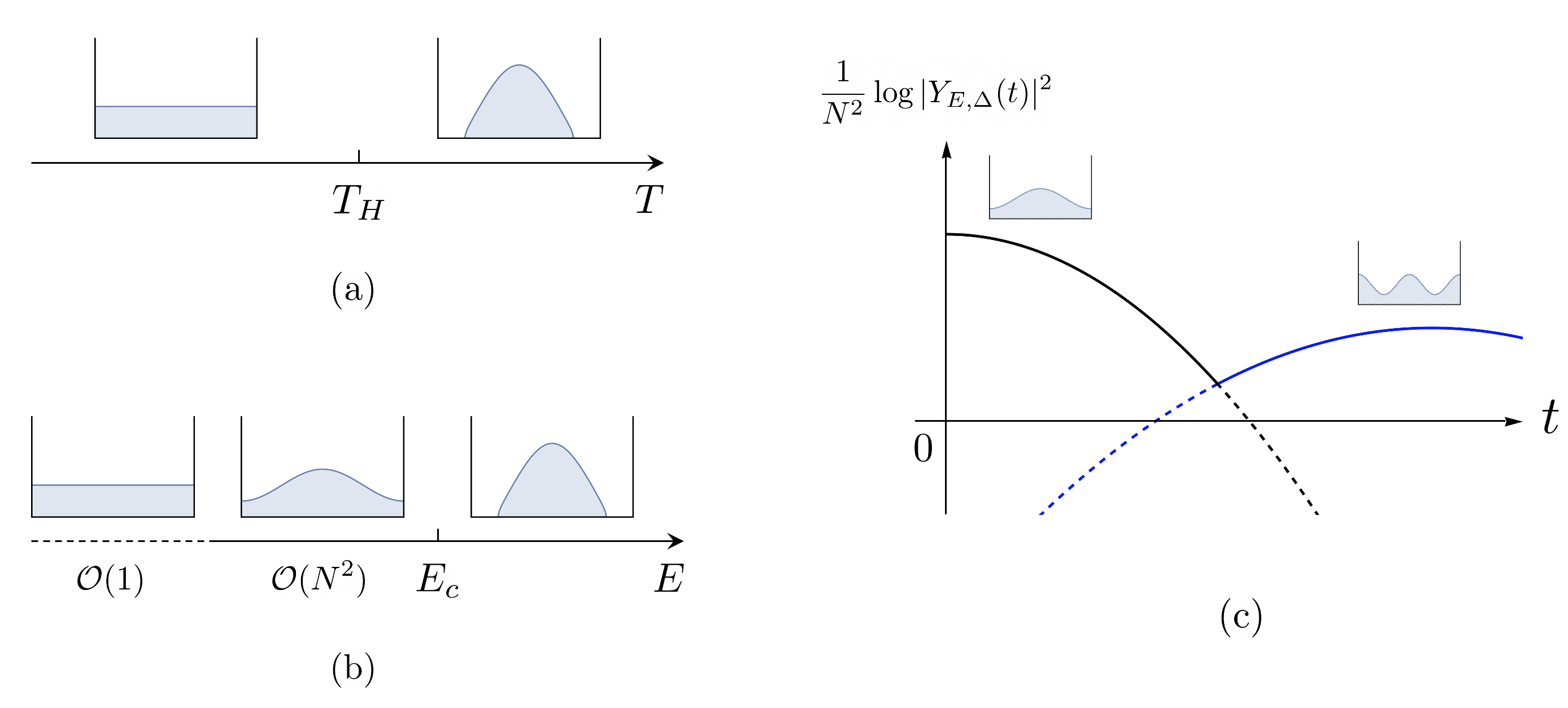}
\caption{(a) The Hagedorn transition in the canonical ensemble and the corresponding eigenvalue distributions. (b) Same transition but in the microcanonical ensemble. The Hagedorn temperature $T_H$ corresponds to a wide range of energy. (c) A snapshot of a possible behavior in the spectral form factor. The early time decay is replaced by a new saddle point.}
\label{fig:gaugeintro}
\end{center}
\end{figure}

Let's briefly discuss the story in the gauge theory case. For an $U(N)$ ($SU(N)$) gauge theory on some compact spatial manifold $\mathcal{M}$, the computation of the thermal partition function can be reduced to a matrix integral over a single $U(N)$ ($SU(N)$) matrix $U$ \cite{Aharony:2003sx}
\begin{equation}\label{Mint}
	Z(\beta) = \int \mathcal{D}U \, e^{-S[U,\beta]}
\end{equation}
where $U$ is the holonomy around the Euclidean time circle, i.e. $U= e^{i\beta \alpha}$ with $\alpha$ being the zero mode of the gauge field in the time direction. The explicit form of the action $S[U,\beta]$ is known for free theories and it only depends on the single particle spectrum on manifold $\mathcal{M}$.\footnote{For readers that are unfamiliar with this topic, we emphasize that (\ref{Mint}) is a matrix integral which produces the exact answer for a single theory at finite $N$, and should not be confused with the use of matrix integral in recent discussions of spectral form factor.} This method has been shown useful in understanding the Hagedorn transition in weakly coupled large $N$ gauge theories. The thermal holonomy has $N$ eigenvalues $\{e^{i\theta_1}, e^{i\theta_2}, ..., e^{i\theta_N}\}$ and in the large $N$ limit one can consider the eigenvalue density $\rho(\theta)$, supported between $[-\pi,\pi]$ and normalized to $1$. For a free theory, there generally is a first order transition at the Hagedorn temperature $T_H = 1/\beta_H$ \cite{Sundborg:1999ue}, below which the large $N$ saddle point $\rho(\theta)=\frac{1}{2\pi}$ is completely uniform, while above the Hagedorn temperature $\rho(\theta)$ starts to localize and opens up a gap, see fig. \ref{fig:gaugeintro} (a). In other words, the Polyakov loop (in the fundamental representation) $\frac{1}{N}\textrm{tr}(U)$ gets a vev above the Hagedorn temperature.\footnote{In a $SU(N)$  theory with only adjoint matter fields, there is a $\mathbb{Z}_N$ center symmetry which multiplies $U$ by a phase $e^{i2\pi k/N},\, k = 0,...,N-1$ (it would be $U(1)$ in a $U(N)$ theory). Correspondingly there are $N$ saddle points above the Hagedorn temperature, after summing over which the expectation of the Polyakov loop will be zero. One can get around this by adding small amount of fundamental matter fields, or by looking at the square of the Polyakov loop. See \cite{Aharony:2003sx} for more details.  } The eigenvalue density $\rho(\theta)$ of the thermal holonomy plays the role of ``geometry" in this problem.

The computation in (\ref{Mint}) can be generalized to $Z(\beta+it)$ \cite{Copetti:2020dil} as well as $Y_{E,\Delta}(t)$. Due to the existence of latent heat in a first order transition, the Hagedorn temperature $T_H$ corresponds to a wide range of energy of order $N^2$ in the microcanonical ensemble. Consider an energy window within this range, the short time saddle of $Y_{E,\Delta}(t)$ is given by a upgapped density $\rho(\theta)$ with a $\cos\theta$ variation, see fig. \ref{fig:gaugeintro} (b).\footnote{See \cite{Hanada:2018zxn,Hanada:2019czd} for discussion on the interpretaion of this saddle in terms of ``partial deconfinement".} This saddle leads to an answer for $|Y_{E,\Delta}(t)|^2$ which decays exponentially with respect to $t^2$. However, one finds new saddle points that start to dominate, which carry higher frequency variations $\cos (n\theta)$, $n>1$. An example will be a saddle point with $\cos 2\theta$ variation, in which $\frac{1}{N}\langle \textrm{tr} (U) \rangle$ is zero while $\frac{1}{N}\langle \textrm{tr} (U^2) \rangle$ becomes nonzero. The new saddles prevent $|Y_{E,\Delta}(t)|^2$ from decaying to exponentially small, which we sketch in fig. \ref{fig:gaugeintro} (c). In section \ref{sec:gauge}, we discuss the new saddles and their properties in detail.

The discussion in perturbative string theory is in many ways analogous to the gauge theory discussion. It is well known that the density of states of a free string gas has a Hagedorn growth, $\rho(E) \sim e^{\beta_H E}$, which can be characterized by the winding modes with winding number $w=\pm 1$ on a thermal manifold $S^1 \times \mathcal{M}$, that are massless at the Hagedorn temperature. The winding modes are the analogs of ``geometries" in this context. If we consider a string gas with energy $E \gg 1/\sqrt{\alpha'}$, the Hagedorn density of states leads to an exponentially decaying $|Y_{E,\Delta}(t)|^2$, much like the black hole or the gauge theory cases. 

\begin{figure}[t!]
\begin{center}
\includegraphics[scale=0.23]{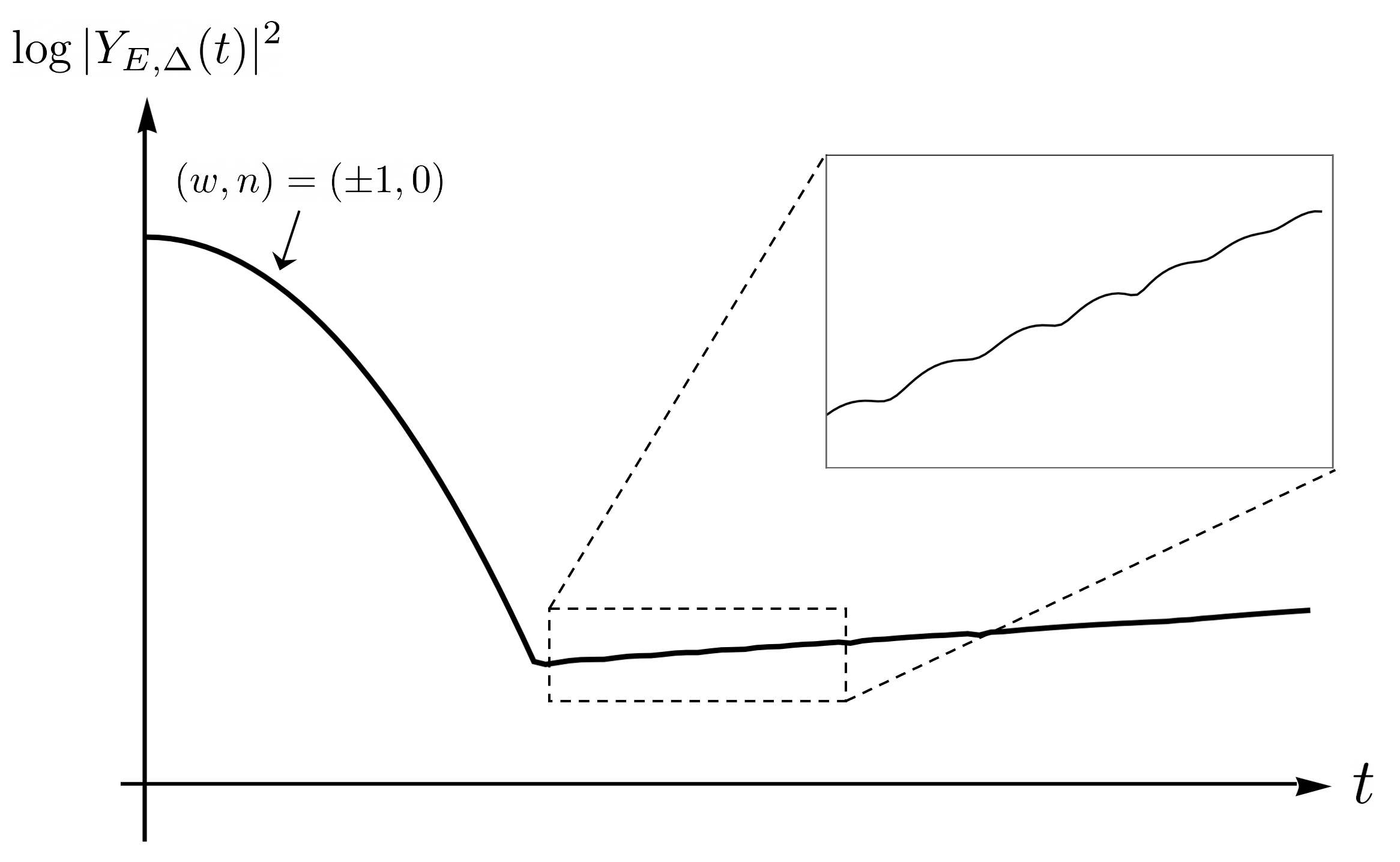}
\caption{The spectral form factor for a free string gas with high energy. The usual winding mode with $w=\pm 1$ and no momentum (in bosonic or Type II theories) gives the early decay. If we zoom in into the latter part, we see bumps and each can be identified with some other winding modes. This graph is in fact generated from numerics, which we discuss in detail in section. \ref{sec:string}.}
\label{fig:stringintro}
\end{center}
\end{figure}

It is important to highlight that once we include the gravity attraction of the string gas, there is a literal connection between winding modes and geometries, due to the ``string star" solution by Horowitz and Polchinski \cite{Horowitz:1997jc}. This is a classical solution much like the Euclidean black hole. It gives the continuous density of states, but it doesn't give a discrete spectrum. Therefore it leads to an exponentially decaying $|Y_{E,\Delta}(t)|^2$. This begs the question ``\emph{what needs to be added to the Horowitz-Polchinski solution in order to get an answer consistent with a discrete spectrum}".\footnote{I'm grateful to Stephen Shenker for asking this question, which prompted some of these considerations.}

Again, new ``geometries" come in, this time being winding modes that carry some other quantum numbers. For both bosonic string and superstrings in flat space, the relevant string modes carry momentum number $n\neq 0 $ along the time direction, as well as some oscillator numbers in the spatial directions. The crucial properties of these string modes is that they become massless at inverse temperatures $\tilde{\beta}$ that are complex, and they each leads to a growing contribution to $|Y_{E,\Delta}(t)|^2$ which peaks at $t=\textrm{Im}(\tilde{\beta})$. We sketch the behavior in fig. \ref{fig:stringintro}.  In section \ref{sec:string}, we present the detailed argument for the statements above.

We are able to understand the contributions of these other string modes explicitly in the free string limit. However, it seems quite natural to speculate that once interaction is included, there could exist a Horowitz-Polchinski like solution for each of the string modes appearing in $|Y_{E,\Delta}(t)|^2$, which one could potentially construct in the canonical ensemble with complex temperature close to the value where the mode becomes massless. This points to the ingredients one needs to add to the Horowitz-Polchinski solution such that the answer respects a discrete spectrum.

As we will discuss, in both stories, the new saddles that appear have their origins as nonperturbatively small corrections to the leading continuous density of states. Their effects are insignificant in the usual thermodynamic discussions, but they gradually expose themselves in $Y_{E,\Delta}(t)$ as we increase $t$.

The rest of the paper is organized as follows. We discuss the free gauge theory and perturbative string theory in section \ref{sec:gauge} and section \ref{sec:string}, respectively. The two sections can be read independently, though they share a lot of similarities. We'll come back to the black hole problem in section \ref{sec:bh} and discuss it in more detail. There we also comment on some potential ways that the discussion in section \ref{sec:gauge} and \ref{sec:string} could be related to black holes. Finally, in section \ref{sec:discussion}, we end with some discussion.

\section{Free large $N$ gauge theory}\label{sec:gauge}

In the following subsections, we discuss the spectral form factor in free large $N$ gauge theories, by doing large $N$ saddle point analysis of a matrix integral of the thermal holonomy. The early time saddle leads to a rapid decay of the spectral form factor. The focus of the discussion will be the appearance of a new family of saddles and their properties.

\subsection{Setting up the calculation}\label{sec:setup}

In this section, we review some basic notions involved in computing the thermal partition function of free gauge theories and generalize them to compute $Y_{E,\Delta}(t)$. 

As was shown in \cite{Aharony:2003sx}, the thermal partition function of free Yang-Mills gauge theory can be written exactly as
\begin{equation}\label{gaugeZ}
	Z(\beta) = \int \mathcal{D}U \, \exp\left\{ \sum_{R} \sum_{m=1}^{\infty}\frac{1}{m} z_{m}^{R}(\beta) \chi_{R}(U^m)\right\},
\end{equation}
where $U$ is an element of the gauge group, $\mathcal{D}U$ is the invariant Haar measure on the group manifold (normalized to one) and $\chi_R$ is the character in representation $R$. $z_{m}^R(\beta)$ is defined as
\begin{equation}\label{BF}
	z_{m}^{R}(\beta) \equiv z_{B}^R(m\beta) + (-1)^{m+1} z_{F}^{R} (m\beta),
\end{equation}
where $z_{B}^R$ and $z_{F}^R$ are the bosonic and fermionic single-particle partition functions for states in representation $R$ (without counting the degeneracy from the size of the representation),
\begin{equation}
	z_{B}^{R} (\beta) = \sum_{R_i = R}e^{-\beta E_i},\quad z_{F}^{R} (\beta) = \sum_{R_i' = R}e^{-\beta E_i'}.
\end{equation}
As mentioned in the introduction, the physical interpretation of $U$ is a Wilson loop (averaged over space) wrapping on the Euclidean thermal circle. The idea behind deriving (\ref{gaugeZ}) is that we can integrate out all the massive modes on the spatial manifold $\mathcal{M}$, leaving an effective action for the zero mode of the gauge field in the time direction.

We will focus on the situation where all the matter fields are in the adjoint representation, $R = \textrm{adj}$, in which case $\chi_{\textrm{adj}}(U^m) = \textrm{tr}(U^m)\textrm{tr}(U^{\dagger m}) $ for $U(N)$ gauge theory and $\chi_{\textrm{adj}}(U^m) = \textrm{tr}(U^m)\textrm{tr}(U^{\dagger m})-1$ for $SU(N)$. We will stick with formulas for $U(N)$ for simplicity, though their difference is negligible in the large $N $ limit. We expect the lessons we will learn should generalize to cases with fundamental matter \cite{Schnitzer:2004qt}, though we did not check this explicitly.

As a concrete example that we will use to demonstrate the idea later, we could consider a pure Yang-Mills theory on $S^3$ with unit radius. In this case, we have
\begin{equation}\label{Zbeta}
	Z(\beta) = \int \mathcal{D}U \, \exp\left\{ \sum_{m=1}^{\infty}\frac{1}{m} z_{m}(\beta) \textrm{tr}(U^m)\textrm{tr}(U^{\dagger m})\right\},
\end{equation}
with
\begin{equation}\label{ex1}
\textrm{Example 1:} \quad	z_{m} (\beta) = z_{B} (m\beta) ,\quad z_{B} (\beta) = \frac{6 e^{-2\beta} - 2 e^{-3\beta}}{(1-e^{-\beta})^3}, \quad \beta_H \approx 1.317.
\end{equation}
In this example $z_{F} (\beta) = 0$ so it is slightly simpler. The discussion can be easily generalized to cases with fermionic excitations as well, as we will discuss briefly later. In (\ref{ex1}) we've also given the inverse Hagedorn temperature in this model, which will be a useful reference later.

The free pure Yang-Mills theory on $S^3$ has the property that all the energy eigenstates are spaced by integers, so $Z(\beta)$ will be a periodic function in the imaginary direction, i.e. $Z(\beta+ 2\pi i) = Z(\beta)$. This might make it seem like that the periodicity is important in some of the following discussion. To disentangle the important physics from the periodicity, we will look at another toy example in which all the states of the system are built with two harmonic oscillators with incommensurable energies $1, \,1/\varphi = 0.618...$, for which $Z(\beta)$ will no longer be periodic. In this case
\begin{equation}\label{ex2}
\textrm{Example 2:} \quad	z_{m} (\beta) = z_{B} (m\beta) ,\quad z_{B} (\beta) = e^{-\beta} + e^{-\beta/\varphi}, \quad \beta_H \approx 0.874.
\end{equation}
This is also the model for which we can compare the analytical results with numerics, as we'll show in fig. \ref{fig:final2}.
As an extreme, we could imagine putting the gauge theory on a manifold with an arbitrary shape, for which the single particle spectrum could start to display chaotic behavior. Of course, even then the model will only have single-particle chaos, not many-body chaos. The spectral form factor in systems with single-body chaos has also been studied in the quadratic SYK model \cite{Winer:2020mdc,Liao:2020lac}.

Finally, let us discuss that how we compute $Y_{E,\Delta} (t)$ from expression (\ref{Zbeta}). The computation involves two steps, the first step is a Hubbard-Stratonovich transformation, which decouples the product $\textrm{tr}(U^m)\textrm{tr}(U^{\dagger m})$ into a sum, at the cost of introducing an extra integral for each $m$,\footnote{More precisely, for each product we need to introduce two integrals. There is another parameter $\tilde{g}_m$ which multiplies $(\textrm{tr}(U^m)- \textrm{tr}(U^{\dagger m}))$. For the saddle points we will discuss, which are symmetric under $U\rightarrow U^\dagger$, $\tilde{g}_m$ can be consistently set to zero. }
\begin{equation}\label{ZHS}
	Z(\beta) =    \int  \prod_{m=1}^\infty dg_m\,g_m \exp\left( - \frac{N^2 m \,g_m^2}{4 z_{m} (\beta)} \right) \int \mathcal{D}U \, \exp\left( \sum_{m=1}^{\infty} \frac{Ng_m}{2} (\textrm{tr}(U^m)+\textrm{tr}(U^{\dagger m}))\right) ,
\end{equation}
where the integration contour of $g_m$ is from the origin to infinity. The detail of this transformation can be found in for example \cite{Klebanov:1994kv,Liu:2004vy,Copetti:2020dil}. The goal of this transformation is solely putting the matrix integral part into a simpler form, for which we already know the phase structure due to \cite{Jurkiewicz:1982iz}. If there were only the $m=1$ term, the matrix integral reduces to the Gross-Witten-Wadia model \cite{Gross:1980he,Wadia:1980cp}. We could pack up the answer of the matrix integral part into a single function $F$ of $g_1, g_2, ... $, and write $Z(\beta)$ as
\begin{equation}\label{ZHS1}
	Z(\beta) =    \int  \prod_{m=1}^\infty dg_m\,g_m \exp\left( - \sum_{m=1}^\infty \frac{N^2 m \,g_m^2}{4 z_{m} (\beta) } - N^2 F(g_1,g_2,...) \right),
\end{equation}
where
\begin{equation}\label{MM}
	\exp\left[-N^2 F(g_1,g_2,...)\right] \equiv 
\int \mathcal{D}U \, \exp\left( \sum_{m=1}^{\infty} \frac{Ng_m}{2} (\textrm{tr}(U^m)+\textrm{tr}(U^{\dagger m}))\right).
\end{equation}

The second step will be computing $Y_{E,\Delta}(t)$ using $Z(\beta)$. This can be done as follows. By definition, the density of states $\rho(E)$ is the inverse Laplace transformation of $Z(\beta)$,
\begin{equation}\label{uprho}
	\rho(E) = \int_{\mathcal{C}} \frac{d\beta}{2\pi i}\, e^{\beta E} Z(\beta).
\end{equation}
where $\mathcal{C}$ is a contour along the imaginary axis, with real part greater than any singularities of $Z(\beta)$. 
(\ref{uprho}), if computed exactly, should give us a sum of delta functions. Now, 
\begin{equation}\label{Yderive}
\begin{aligned}
	Y_{E,\Delta}(t) & = \int d\tilde{E}\, e^{- \frac{(\tilde{E} - E)^2}{2\Delta^2}} e^{-i\tilde{E} t} \rho(\tilde{E}) = \int d\tilde{E}\, e^{- \frac{(\tilde{E} - E)^2}{2\Delta^2}} e^{-i\tilde{E} t}\int_{\mathcal{C}} \frac{d\beta}{2\pi i}\, e^{\beta \tilde{E}} Z(\beta)\\
	& = \int_{\mathcal{C}} d\beta \frac{ \Delta }{\sqrt{2\pi} i}\, e^{(\beta - it) E + \frac{1}{2}(\beta - it)^2 \Delta^2} Z(\beta) \\
	& = \int_{\mathcal{C}} d\beta \frac{ \Delta }{\sqrt{2\pi} i}\, e^{ \beta E + \frac{1}{2}\beta^2 \Delta^2} Z(\beta +it).
\end{aligned}
\end{equation}
The last line is the expression that appears in \cite{Saad:2018bqo}. Finally, we can plug (\ref{ZHS1}) into (\ref{Yderive}) and get
\begin{equation}\label{Yfinal}
	Y_{E,\Delta}(t) = \int_{\mathcal{C}}  \frac{ \Delta \,d\beta }{\sqrt{2\pi} i}\int  \prod_{m=1}^\infty dg_m \,g_m     \exp\left[ \beta E + \frac{1}{2}\beta^2 \Delta^2  - \sum_{m=1}^{\infty} \frac{ N^2 m \,g_m^2}{4 z_{m} (\beta + it) } - N^2 F(g_1,g_2,...) \right].
\end{equation}
This is the main result of this section and will be the starting point of our computation. We will look for large $N$ saddle points of this quantity. Despite its complicated form, the actual saddle points we will identify in the next few sections will be quite simple.

\subsection{The saddle points in the microcanonical ensemble}\label{sec:micro}

In this section, we discuss the saddle points that arise when we consider the microcanonical ensemble,\footnote{Usually microcanonical ensemble is defined with a hard cutoff, while $Y_{E,\Delta}$ has a Gaussian filter in it, but we stick to this terminology nonetheless. The benefit of a Gaussian cutoff, apart from being easier to compute, is to get rid of oscillations in $|Y_{E,\Delta}(t)|^2$ coming from the hard cutoff \cite{Gharibyan:2018jrp}. We should note that the average energy in $Y_{E,\Delta}(t=0)$ is generally not $E$ but rather $E+\Delta^2 S'(E)$.} namely $Y_{E,\Delta}(t)$ at $t=0$.

Let's first briefly review how to derive the standard story in the canonical ensemble, by looking at $Z(\beta)$ in (\ref{ZHS1}). In this case there are two relevant phases of the matrix integral (\ref{MM}). The first is the ungapped phase\footnote{``gapped", ``ungapped" refer to the property of the eigenvalue density $\rho(\theta)$ of thermal holonomy. They have nothing to do with the physical spectrum. } in which the large $N$ saddle point of the density of eigenvalues $\rho(\theta)$ is given by
\begin{equation}\label{rhotheta}
	\rho(\theta) = \frac{1}{2\pi} \left(1 + \sum_{m=1}^{\infty} m g_m  \cos(m\theta)\right),
\end{equation}
for which in the large $N$ limit
\begin{equation}\label{Fungap}
	F(g_1,g_2,...) = - \sum_{m=1}^{\infty} \frac{m}{4} g_m^2 , \quad \quad\textrm{(ungapped phase).}
\end{equation}
Plugging (\ref{Fungap}) into (\ref{ZHS1}) we get
\begin{equation}\label{Zungap}
	Z(\beta) =    \int  \prod_{m=1}^\infty dg_m\,g_m \exp\left( - \sum_{m=1}^\infty \frac{N^2 m \,g_m^2}{4} \left(\frac{1}{z_{m}(\beta)} - 1\right) \right).
\end{equation}
When the temperature is low, or $\beta\gg 1$ , we have $z_m(\beta) <1$ for all $m$, which can be seen in the examples (\ref{ex1}) and (\ref{ex2}). Therefore the saddle point of (\ref{Zungap}) will be at $g_1 = g_2 =... = 0$. This leads to a completely uniform $\rho(\theta$) with no variation and therefore the expectation value of the Polyakov loop $\frac{1}{N} \langle\textrm{tr}(U)\rangle$ is zero in this phase, as well as the higher moments $\frac{1}{N}\langle\textrm{tr}(U^m)\rangle =  0$.

\begin{figure}[t!]
\begin{center}
\includegraphics[scale=0.33]{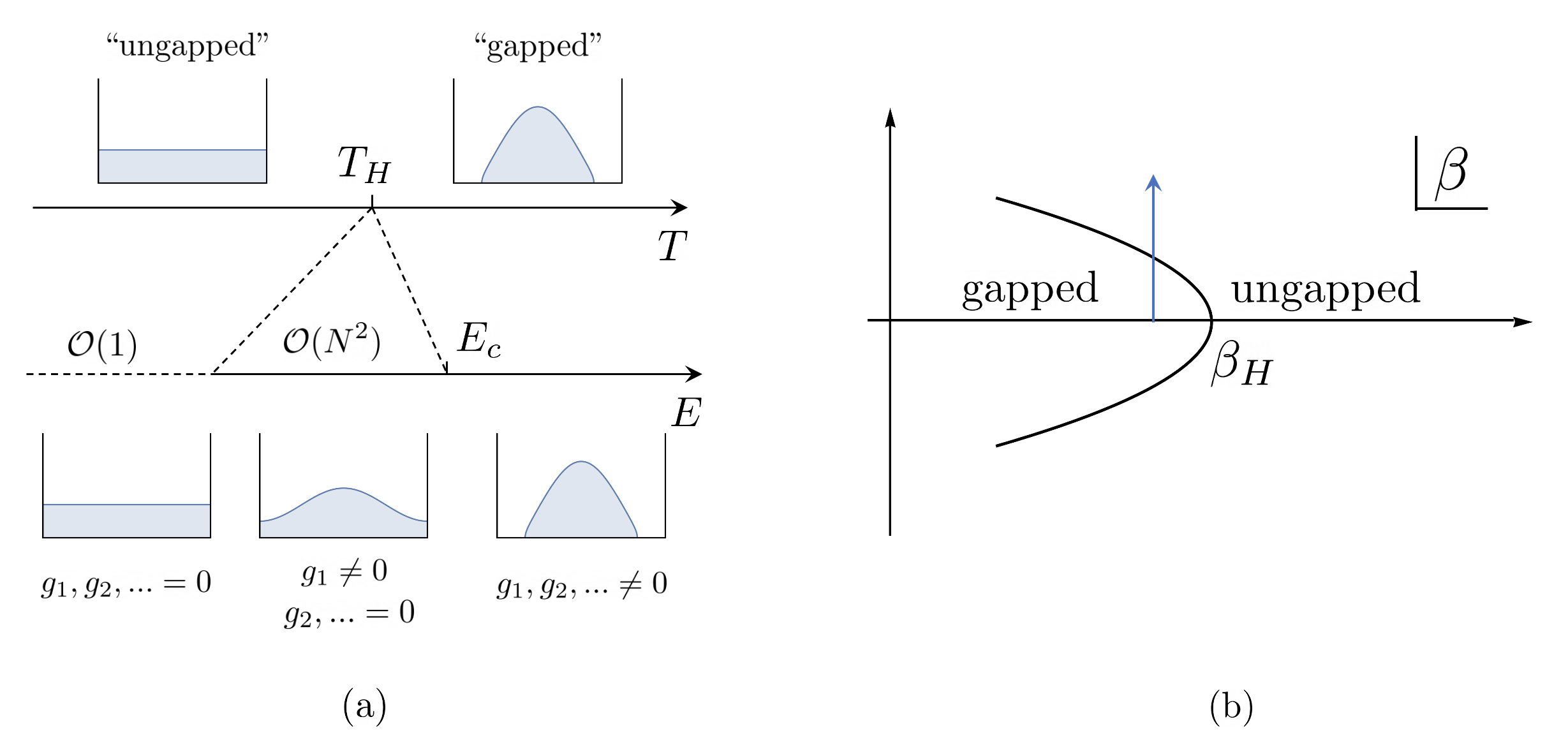}
\caption{(a) The phase structure in the canonical ensemble (top) and the microcaonincal ensemble (down) at $t=0$. (b) If we consider the quantity $Z(\beta+it)$ and increase $t$ (along the blue arrow), we will soon encounter a phase transition and go into the ungapped phase.}
\label{fig:phasegauge}
\end{center}
\end{figure}

As we increase the temperature, or decrease $\beta$, $z_m(\beta)$ increase and $z_1(\beta)$ is the first to reach one. The real solution of $z_1(\beta) = 1$ is denoted as $\beta_H = 1/T_H$, the inverse Hagedorn temperature. When $z_1(\beta)=1$, $g_1$ becomes tachyonic in (\ref{Zungap}), signaling a phase transition, i.e. the deconfinement transition, which is of first order. It turns out that when $\beta<\beta_H$, the relevant phase of the matrix integral (\ref{MM}) becomes a one-cut gapped phase. The exact solution of the matrix integral is slightly more involved, but can be found in \cite{Jurkiewicz:1982iz} as well as \cite{Aharony:2003sx}. Here we will not write down the explicit expression for $\rho(\theta)$ and $F(g_1,g_2,...)$, but the conclusion is that in this phase we have all the coefficients $g_1,g_2,...\neq 0$. All the moments of the thermal holonomy have nonzero expectation values, $\frac{1}{N}\langle\textrm{tr}(U^m)\rangle\neq 0$. The summary of the phase structure in the canonical ensemble can be seen in the top panel of fig. \ref{fig:phasegauge} (a).

In this model, if we consider the quantity $Z(\beta+it)$, starting at the gapped phase at $\beta < \beta_H$ and increase $t$, one will transit into the ungapped phase. Or, if one starts in the ungapped phase, at some finite and fixed $t$, one needs to go to a higher temperature to reach the deconfined phase. This is illustrated in fig. \ref{fig:phasegauge} (b). This phenomena is termed ``delayed deconfinement transition" and was studied in detail in \cite{Copetti:2020dil}.\footnote{As a side comment, we think that the one dimensional transition line in fig. \ref{fig:phasegauge} (b) can be understood as the condensation of Lee-Yang zeros \cite{PhysRev.87.404} on the complex $\beta$ plane at large $N$. See \cite{Kristensson:2020nly} for a related discussion.} This is analogues to the transition from black hole to thermal AdS we mentioned in the introduction, and for this reason, the quantity $Z(\beta+it)$ is not suitable to study the long time behavior of high energy states. For that we need to consider the microcanonical ensemble quantity $Y_{E,\Delta}(t)$, which we now come to. 

We first discuss what are the saddles that appear in $Y_{E,\Delta}(t=0)$, as we vary $E$. This is simply the microcanonical ensemble, with the only difference being that there is a Gaussian window in defining $Y_{E,\Delta}$. Therefore we can basically translate the canonical ensemble answer into the microcanonical one, namely that at low energies, the saddle point is completely uniform, while at sufficiently high energy, the saddle point is gapped. However, we should note that in a first order transition, there is a latent heat, namely the transition temperature $T_H$ in the canonical ensemble in fact corresponds to an intermediate range of energy, which is of order $N^2$ in the microcanonical ensemble. In this range, the eigenvalue distribution $\rho(\theta)$ is ungapped but is also not uniform. To see this, we can take (\ref{Fungap}) into (\ref{Yfinal})
 \begin{equation}\label{Ynew}
	Y_{E,\Delta}(t=0) = \int_{\mathcal{C}}  \frac{ \Delta \,d\beta }{\sqrt{2\pi} i}\int  \prod_{m=1}^\infty dg_m \,g_m     \exp\left[ \beta E + \frac{1}{2}\beta^2 \Delta^2 - \sum_{m=1}^\infty \frac{N^2 m \,g_m^2}{4} \left(\frac{1}{z_{m}(\beta)} - 1\right)  \right]
\end{equation}
and look for saddle points for $\beta,g_1,g_2,...$. The equation for $\beta$ is
\begin{equation}\label{eq1}
	E + \beta \Delta^2 + \sum_{m=1}^\infty \frac{N^2 m \,g_m^2}{4} \frac{z_m'(\beta )}{z_{m}(\beta )^2}  = 0,
\end{equation}
and the equations for $g_m, m =1,2,...$ are
\begin{equation}\label{eq2}
	g_m \left(\frac{1}{z_{m}(\beta )} - 1\right) = 0,\quad m=1,2,...\,.
\end{equation}
From (\ref{eq2}) we see that either $g_m = 0$ or $z_m(\beta) = 1$. From (\ref{eq1}), we can't have all the $g_m$ being zero (assuming $E\sim \mathcal{O}(N^2)$). Say $g_n \neq 0$, then $\beta$ must solve $z_n(\beta) = 1$. Since the zeros of functions $\{z_m(\beta)-1\}$ don't coincide, all the other $g_m, m\neq n$ must be zero. For now, let's restrict our attention to real $\beta$, so there is a unique real solution of $z_n(\beta) = 1$, denoted by $\beta_n$. From (\ref{BF}) we see that in the cases without fermions, we simply have $\beta_n = \beta_H/n$. The end conclusion is that (\ref{eq1}) and (\ref{eq2}) admit an infinite set of solutions $\{\beta_n,g_1^n, g_2^n ,...\}$ labeled by $n$
\begin{equation}\label{solnn}
	\beta_n = \frac{\beta_H}{n}, \quad g_{m}^n =  \delta_{m}^n  \sqrt{ \frac{E+\beta_H \Delta^2 /n}{N^2} \left( \frac{-4 }{n z_n'(\beta_H/n) } \right)}, \quad  \quad n \in \mathbb{Z}_+.
\end{equation}
The $n$-th solution leads to 
\begin{equation}
	Y_{E,\Delta}(t=0) \sim \exp\left( \frac{\beta_H E}{n} + \frac{\beta_H^2 \Delta^2}{2n^2} \right),
\end{equation}
so the solutions with $n>1$ leads to an exponentially suppressed answer compared with the $n=1$ solution. This is consistent with the fact that all the $n>1$ solutions lead to negative directions of integral for $g_{m<n}$ in (\ref{Ynew}), assuming the integral is along the real axis. Similar unstable solutions were also studied in \cite{Azuma:2012uc}. Here we will not be interested in these solutions further, but turn our focus to the solution with $n=1$:
\begin{equation}\label{n=1}
	\beta=\beta_H, \quad g_1 = \sqrt{ \frac{E+\beta_H \Delta^2 }{N^2} \left( \frac{-4 }{z_1'(\beta_H) } \right)},\quad g_2=g_3 = ... = 0.
\end{equation}
This solution is only applicable when $g_1<1$, since beyond that the density $\rho(\theta)$ (\ref{rhotheta}) will no longer be ungapped and above analysis cease to apply any more. From (\ref{n=1}), we see that the above solution is applicable when the energy window is lower than a critical energy $E_c$\footnote{Due to the fast growing Hagedorn density of state, the average energy in the quantity $Y_{E,\Delta}(t=0)$ is not $E$, but rather $E+\beta_H \Delta^2$.}
\begin{equation}\label{enerrange}
	E + \beta_H \Delta^2 <E_c \equiv N^2 \frac{-z'(\beta_H)}{4}
\end{equation}
beyond which the saddle point will be given by the gapped solution as in the high temperature phase. This slightly lengthy discussion is summarized into a simple phase diagram as in the lower panel of fig. \ref{fig:phasegauge} (a). Let us also record the value of $E_c$ for the two examples we consider. In the example (\ref{ex1}), we have $E_c = 3N^2/4$; in example (\ref{ex2}), we have $E_c = 0.194 N^2$.

The final answer for $Y_{E,\Delta}(t=0)$ for energy below $E_c$ is
\begin{equation}
	Y_{E,\Delta} (t=0)\sim \exp\left( \beta_H E + \frac{\beta_H^2 \Delta^2}{2} \right).
\end{equation}
We note this is what we would have expected if we compute $Y_{E,\Delta}$ using the Hagedorn density of states $\rho(E)\sim e^{\beta_HE}$. We were merely reproducing it from the matrix integral to gain some familiarity.

We should explain why we bother talking so much about the saddle point for intermediate energies. We could have started at a higher energy $E + \beta_H \Delta^2>E_c$, for which the saddle point will be the gapped $\rho(\theta)$, and consider $Y_{E,\Delta}(t)$ of that. The main reason is that the analysis in the gapped phase is mathematically more complicated, while the ungapped phase is much simpler. So it is easier to work with the intermediate range of energy first. After we understand the basic structure, it would be easier to generalize to higher energies, which we do in section \ref{sec:gapped}.

\subsection{$Y_{E,\Delta}(t)$ of the early time saddle and the puzzle}

\begin{figure}[t!]
\begin{center}
\includegraphics[scale=0.3]{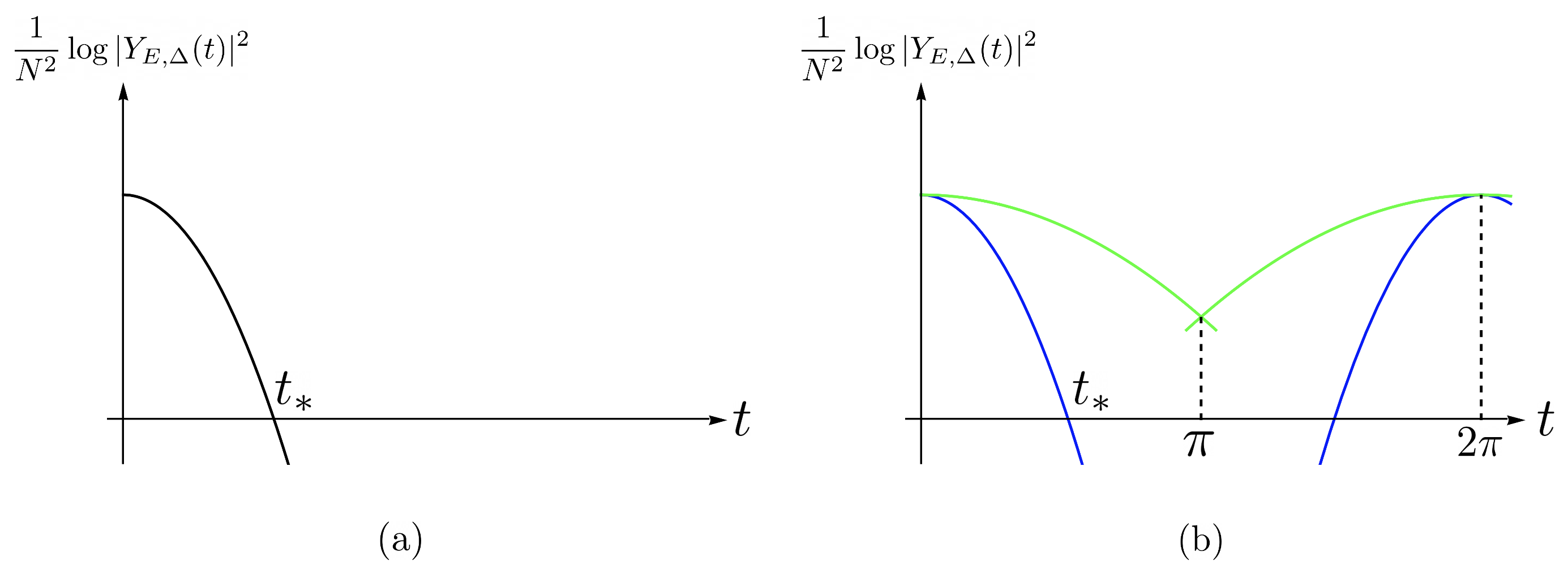}
\caption{(a) $|Y_{E,\Delta}(t)|^2$ from the short time saddle. It decays to exponentially small at $t_* \sim \mathcal{O}(1)$. (b) If the single particle spectrum is evenly spaced (as in our first example (\ref{ex1})), there is another obvious saddle by shifting the short time one by the period. This could sometimes prevent $|Y_{E,\Delta}(t)|^2$ from becoming exponentially small (see green curve), but not for large $\Delta^2$ (see blue curve).}
\label{fig:short}
\end{center}
\end{figure}

After discussing the saddle points for $Y_{E,\Delta}(t)$ at $t=0$, it is simple to track its time dependence. We will focus on energies belong to (\ref{enerrange}). For completeness, let's write down the expression for $Y_{E,\Delta}(t)$ 
\begin{equation}\label{Ynew1}
	Y_{E,\Delta}(t) = \int_{\mathcal{C}}  \frac{ \Delta \,d\beta }{\sqrt{2\pi} i}\int  \prod_{m=1}^\infty dg_m \,g_m     \exp\left[ \beta E + \frac{1}{2}\beta^2 \Delta^2 - \sum_{m=1}^\infty \frac{N^2 m \,g_m^2}{4} \left(\frac{1}{z_{m}(\beta +it)} - 1\right)  \right]
\end{equation}
and the saddle point equations
\begin{equation}\label{eqbeta}
	E + \beta \Delta^2 + \sum_{m=1}^\infty \frac{N^2 m \,g_m^2}{4} \frac{z_m'(\beta + it )}{z_{m}(\beta +it)^2}  = 0,
\end{equation}
\begin{equation}\label{eqg}
		g_m \left(\frac{1}{z_{m}(\beta + it )} - 1\right) = 0,\quad m=1,2,...\,.
\end{equation}
It is easy to see that the solution (\ref{n=1}) at $t=0$ continues into
\begin{equation}\label{early}
	\beta = \beta_H - it, \quad  g_1 =\sqrt{ \frac{E+ (\beta_H -it)\Delta^2 }{N^2} \left( \frac{-4 }{z_1'(\beta_H) } \right)}, \quad g_2 = g_3 = ... = 0.
\end{equation}
This is the early time saddle.\footnote{One might be worried about that $g_1$ becomes complex and some other phases of the matrix model might be important. This is discussed in detail in \cite{Copetti:2020dil}. In general one might need to consider them. However, we checked that generally they alone don't seem to solve the problem we will mention.} The early time saddle gives 
\begin{equation}
	Y_{E,\Delta}(t) \sim \exp\left[ (\beta_H -it) E   + \frac{(\beta_H -it)^2}{2} \Delta^2 \right],
\end{equation}
so the spectral form factor is rapidly decaying
\begin{equation}\label{earlyres}
	|Y_{E,\Delta}(t)|^2 \sim \exp\left[2\beta_H E + (\beta_H^2 - t^2) \Delta^2\right].
\end{equation}
Again, we would have arrived at the same answer just by using the continuous Hagedorn density of states $\rho(E) \sim e^{\beta_H E}$. We sketch the answer given by (\ref{earlyres}) in fig. \ref{fig:short} (a). However, (\ref{earlyres}) shows that the spectral form factor decays to exponentially small, of order $e^{-N^2}$, at time
\begin{equation}\label{smalltime}
	t_* \sim \sqrt{\frac{2\beta_H E + \beta_H^2 \Delta^2}{\Delta^2}} \sim \mathcal{O}(1).
\end{equation}
Note that we are choosing the width $\Delta$ to scale linearly with $N$, so $E$ and $\Delta^2$ are both of order $N^2$. By choosing $\Delta^2/E$ to be a large order one number, we can make $t_*$ in (\ref{smalltime}) as short as $\beta_H$. 

We emphasize that it would be very surprising that $|Y_{E,\Delta}(t)|^2$ decays to $\mathcal{O}(e^{-N^2})$. We recall that $Y_{E,\Delta}(t)$ is a sum of $\mathcal{O}(e^{N^2})$ many order one terms and it would require a delicate cancellation for the final sum to be $\mathcal{O}(e^{-N^2})$. As we will see, this doesn't really happen.

It might be obvious that there could be a simple way out if we were considering our first example in (\ref{ex1}), a pure Yang-Mills theory on $S^3$ with unit radius. The reason is that in this example $Y_{E,\Delta}(t)$ has to be a periodic function with period $2\pi$, as all the states have integral energies. This periodicity is reflected explicitly in (\ref{Ynew1}) as all the $z_m$ are invariant under $t \rightarrow t+2\pi n,\,n\in \mathbb{Z}$, which can be seen from eqn (\ref{ex1}). The direct consequence of this is that the same early time saddle will repeat itself around $t=2\pi n$. If one chooses a small $\Delta^2$, with which (\ref{earlyres}) decays slower, then the repetitions the early time saddle can already prevent  $|Y_{E,\Delta}(t)|^2$ from becoming exponentially small, see the green curve in fig. \ref{fig:short} (b).

However, this does not solve the problem completely. The reason is that with a larger $\Delta^2$ we can achieve $t_* \approx \beta_H \approx 1.317$ in this example. Since $\beta_H < \pi$, we can have a situation as the blue curve in fig. \ref{fig:short} (b), where there is still a gap between the two saddles in which $|Y_{E,\Delta}(t)|^2$ becomes exponentially small. Also, if we consider different examples in which $Y_{E,\Delta}(t)$ does not have a period, then this argument will not apply.

\subsection{A family of new saddles}\label{sec:newsaddle}

In this section, we discuss a family of new saddles that appear and prevent $|Y_{E,\Delta}(t)|^2$ from becoming exponentially small. In the ungapped phase we are considering, they can be characterized by an eigenvalue distribution $\rho(\theta)$ which only has $\cos n \theta$ variation, with $n\in \mathbb{Z}_+$. However, a better way to characterize them, which will also apply to the gapped phase, is that each of them preserve a subgroup $\mathbb{Z}_n$ of the center symmetry.

\begin{figure}[t!]
\begin{center}
\includegraphics[scale=0.35]{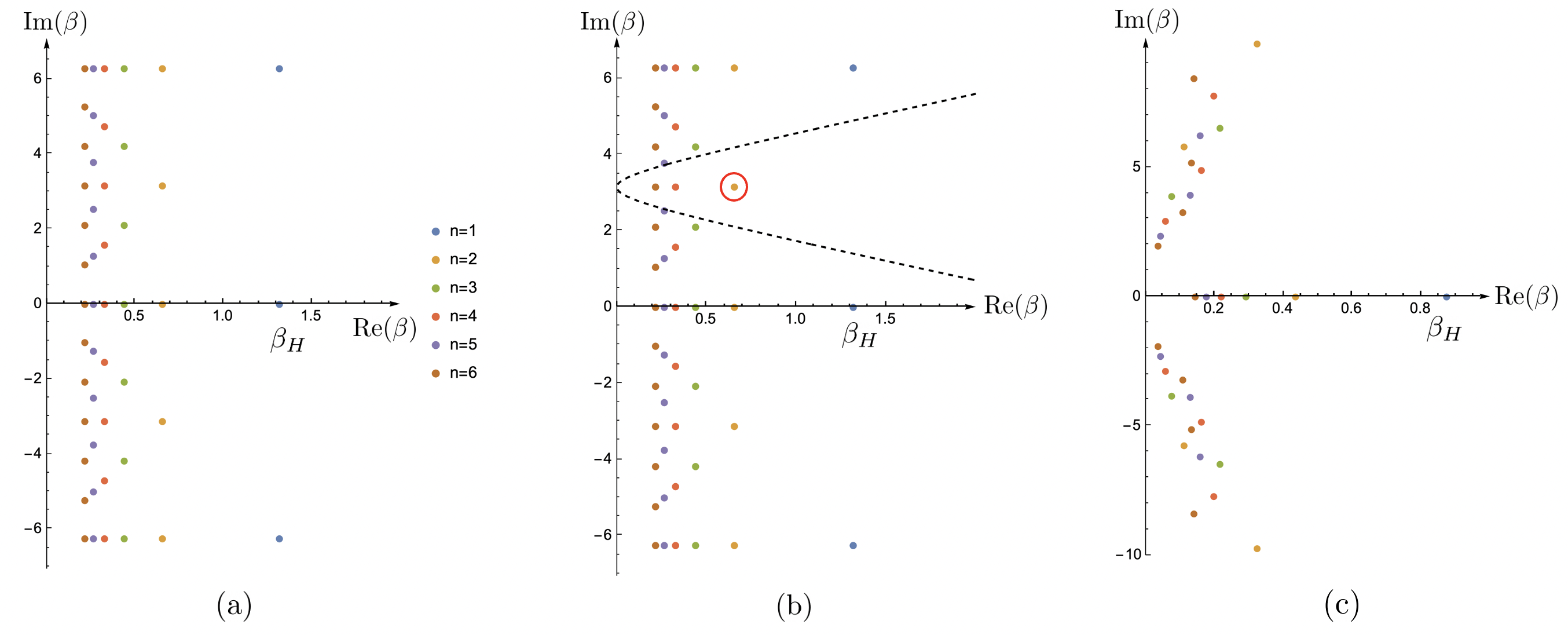}
\caption{(a) The distribution of the zeros for the functions  $\{ z_n(\beta)-1 | n \leq 6\}$, for the example in (\ref{ex1}). The zeros have a periodic structure since all the energy eigenvalues are integers. (b) The saddle point that is most important at time $t$ (circled by red) corresponds to the zero rightmost to the hyperbola (\ref{hyperbola}), shown in dashed line. (c) The zeros for the example in (\ref{ex2}). Unlike (a), the zeros now display a bit of randomness.}
\label{fig:zeros}
\end{center}
\end{figure}

Let's discuss how to see these saddles from the equations (\ref{eqbeta}) and (\ref{eqg}). The saddle points are in fact simple generalizations of those already described in (\ref{solnn}). From (\ref{eqg}), the saddle point value $\beta = \tilde{\beta}$ solves 
\begin{equation}
	z_{n} (\tilde{\beta} +it) = 1,  \quad n\in \mathbb{Z}_+
\end{equation}
and other than $g_n$ all the other $g_{m\neq n}$ are zero. The value of $g_n$ can then be solved from (\ref{eqbeta})
\begin{equation}
	g_n =  \sqrt{ \frac{E+ \tilde{\beta} \Delta^2 }{N^2} \left( \frac{-4 }{n z_n'(\tilde{\beta}+it) } \right)}.
\end{equation}
We could then classify all the saddle points by first finding all the zeros of the functions $\{ z_n(\beta)-1\}$, and the saddle point $\tilde{\beta}$ are given by  subtracting from the zeros by $it$. The crucial difference from the discussion around (\ref{solnn}) is that we should consider zeros that are complex. In fig. \ref{fig:zeros} (a) and (c) we show the distribution of the zeros on the complex plane for the two examples we consider.  

For a saddle point corresponding to a solution of $z_n (\beta) = 1$, it will have a $\cos n \theta$ variation and preserves the $\mathbb{Z}_n$ subgroup of center symmetry.

Now, which are the saddle points we really care about? We are interested in the ones that lead to a large contribution to $|Y_{E,\Delta}(t)|^2$. Since the saddle point with $\tilde{\beta}$ contributes 
\begin{equation}
	|Y_{E,\Delta}(t)|^2 \sim \exp\left( 2\tilde{\beta} E+\tilde{\beta}^2 \Delta^2\right), 
\end{equation}
we see that the requirement for the saddle point to contribute $\mathcal{O}(e^{ N^2})$ is
\begin{equation}
	\textrm{Re} \left(2 \tilde{\beta} E +\tilde{\beta}^2 \Delta^2 \right) > 0.
\end{equation}
Translated into fig. \ref{fig:zeros}, this corresponds to the zeros on the complex $\beta$ plane that are to the right of the hyperbola
\begin{equation}\label{hyperbola}
	\left(\textrm{Re}(\beta) + \frac{E}{\Delta^2} \right)^2  - 	\left(\textrm{Im}(\beta) - t\right)^2 = \frac{E^2}{\Delta^4}.
\end{equation}
We draw one such hyperbola in fig. \ref{fig:zeros} (b). The dominant saddle point is the one that is furthest to the right from the hyperbola, which we mark by a red circle in the example in fig. \ref{fig:zeros} (b). As we increase $t$, the hyperbola shifts upwards on the complex plane, picking up different dominant contributions to the spectral form factor. Varying $E/\Delta^2$ changes the shape of the hyperbola, and depending on it, the final series of dominating saddles can be different. In the two examples we consider (fig. \ref{fig:zeros} (a) and (c)), it is clear that there are always zeros that lie to the right of the hyperbola, so $|Y_{E,\Delta}(t)|^2$ never decays to exponentially small.\footnote{We've not attempted to prove this for an arbitrary single particle partition function $z(\beta)$, but we expect it to be true.}

\begin{figure}[t!]
\begin{center}
\includegraphics[scale=0.35]{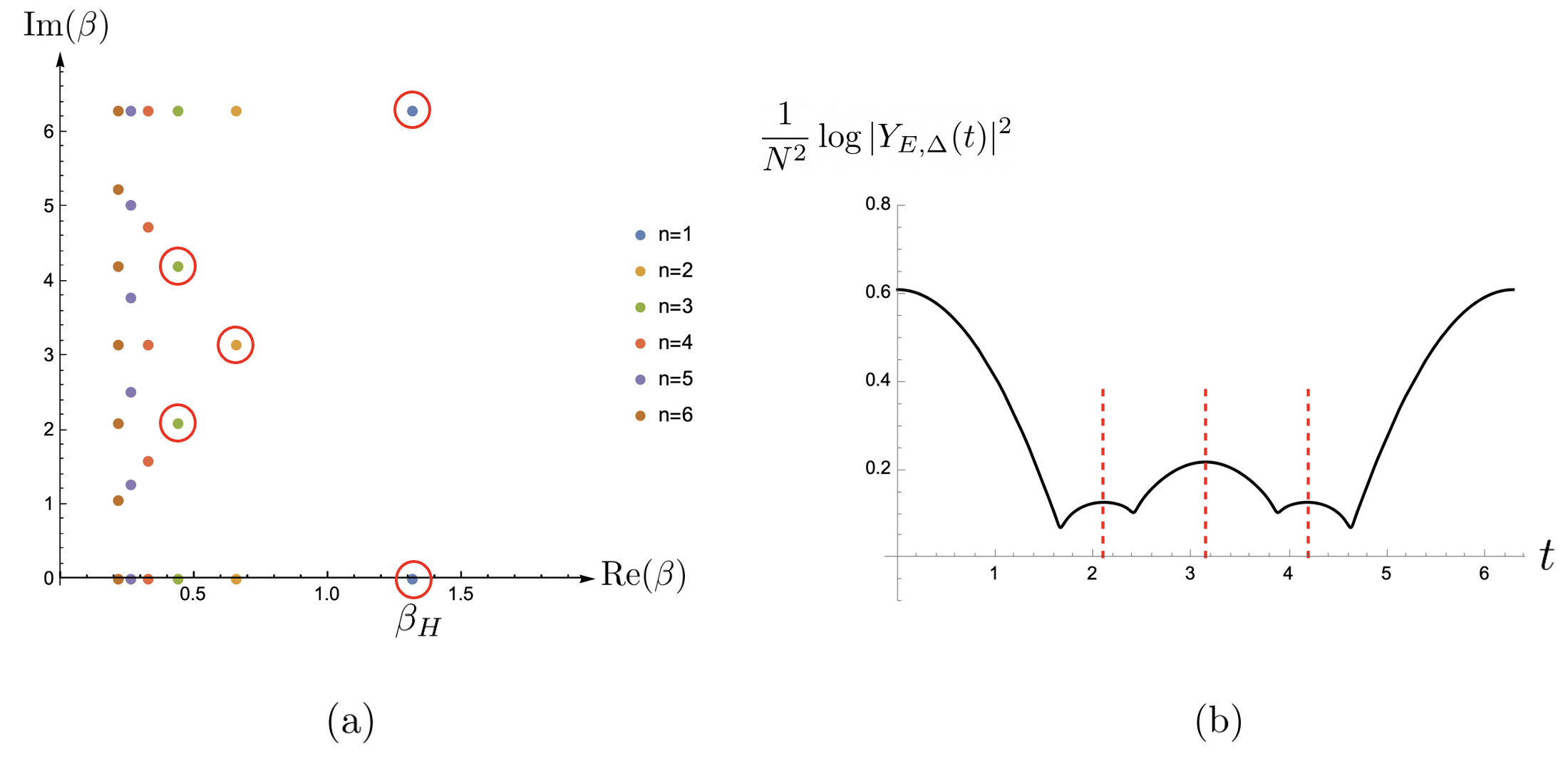}
\caption{(a) In the first example (\ref{ex1}) with parameters $E=0.1N^2, \Delta^2 = 0.2N^2$, we circle out the zeros for which the corresponding saddles will dominate. (b) We plot the analytic prediction for the spectral form factor. The red dashed lines are located at the imaginary parts of the circled zeros in (a).}
\label{fig:final}
\end{center}
\end{figure}

\begin{figure}[t!]
\begin{center}
\includegraphics[scale=0.31]{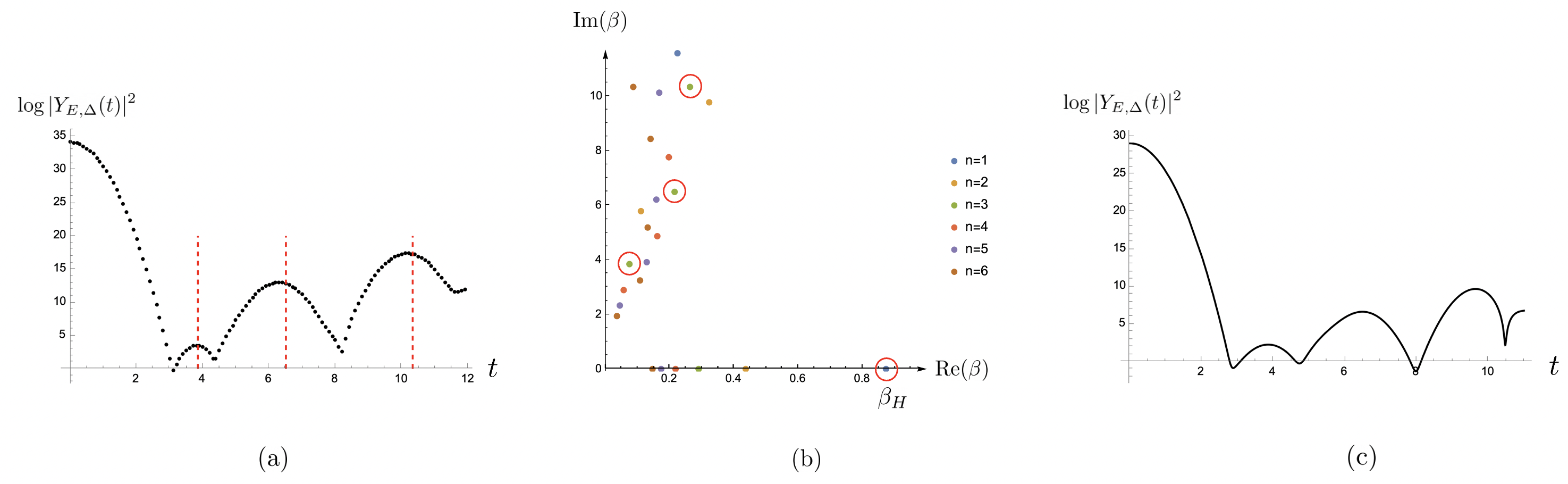}
\caption{(a) We show the numerical answer of the spectral form factor for the example (\ref{ex2}), with $E=15,\Delta^2 =15/4$. (b) We circle out the zeros which correspond to the location of the peaks in (a). Their locations are represented by the red dashed lines in (a). We see that they correspond to $n\leq 3$. We are not seeing the contributions from some zeros with $n>3$ in the numerics. We expect this is a finite $N$ effect ($N=7$). (c) We plot the analytic large $N$ answer if we were to \emph{only} include the contributions from the zeros with $n\leq 3$. We see that it resembles the finite $N$ answer in (a).}
\label{fig:final2}
\end{center}
\end{figure}

In fig. \ref{fig:final}, we circle out the contributing saddles in our first example (\ref{ex1}) and show the final answer for the spectral form factor based on this discussion. In fig. \ref{fig:final2}, we plot the numerical answer of $|Y_{E,\Delta}(t)|^2$ for our second example (\ref{ex2}), computed using the exact spectrum in the case of $N=7$, which can be extracted from a mathematica file accompanying the reference \cite{Kristensson:2020nly}. Presumably due to finite $N$ effect, we are only seeing the contributions from the saddle points with $n\leq 3$ in the numerics.

One might be concerned whether these saddle points contain negative modes. Naively it might seem they do, since other than the early time saddle and its periodic repetitions, these saddles excite higher moments of the thermal holonomy, which seems to naively suggest that the first moment (or its corresponding amplitude $g_1$) is tachyonic. Indeed we've argued in section \ref{sec:micro} that this happens for the saddles that correspond to zeros on the real axis of fig. \ref{fig:zeros} (a), (c) (other than $\beta_H$).  However, as we can explicitly check, this isn't necessarily true for zeros that are on the complex plane. For example, consider our first example (\ref{ex1}) and the saddle point with $\tilde{\beta} + it = \beta_H/2 + i\pi$, we have
\begin{equation}
	z_{1} (\tilde{\beta} + it) = z_{1} \left(\frac{\beta_H}{2} + i\pi \right) \approx 0.54 <1,
\end{equation}
so the amplitude $g_1$ is actually massive at this point. In general, for zeros that are on the complex plane, the masses for $g_m$ are complex, so one needs to be more careful about the integration contour of $g_m$ in order to decide whether the solution has a negative direction. Of course, even if it has a negative direction, it won't be a major issue for our current discussion here, since it would likely mean that there is an even more dominant solution, giving a larger contribution to spectral form factor. 

Before we end this section, let's make some comments on the saddles we identify. These saddles can be easily distinguished via order parameters $\frac{1}{N} \langle \textrm{tr}(U^n)\rangle, n \in \mathbb{Z}_+$. A better way to characterize them is based on the subgroup $\mathbb{Z}_n$ of the center symmetry they preserve. This also applies to the gapped phase, which we will discuss with some preliminary results in section \ref{sec:gapped}. We don't know whether there is any \emph{Lorentzian} meaning of these different phases.\footnote{See \cite{Unsal:2010qh,Betzios:2017yms,Cherman:2020zea} for some other cases where similar type of solutions can arise, where the Lorentzian meaning might be clearer.}

These saddles can be vaguely interpreted as coming from non-perturbatively small corrections to the continuous density of states $\rho(E)$. To see this, we note that if we extrapolate their contributions to $Y_{E,\Delta}(t)$ to $t=0$, they each gives a non-perturbatively small contribution compared to the leading answer. However, we used the word ``vaguely" because generally it is unclear whether these saddle points still contribute at $t=0$. Note that these saddle points are real when $\tilde{\beta}$ is real, that is, when the time $t$ aligns with the imaginary part of the complex zero, while in general they are complex.

We also note that even though we were considering $E\sim \mathcal{O}(N^2)$ above, the structure we saw can be traced back to lower energies, $E\sim \mathcal{O}(1)$, or below the Hagedorn temperature. To see this, we note that the partition function below the Hagedorn temperature has the following form in the infinite $N$ limit (up to normalization) \cite{Sundborg:1999ue,Polyakov:2001af,Aharony:2003sx}
\begin{equation}\label{ZNinfty}
	Z_{N = \infty}(\beta) = \prod_{m=1}^\infty \frac{1}{1-z_{m}(\beta)},
\end{equation}
therefore the zeros shown in fig. \ref{fig:zeros} are simply the singularities of $Z_{N = \infty}(\beta)$ on the complex plane. Therefore, each of the saddle points we identify has its root in a complex singularity of  $Z_{N = \infty}(\beta)$. This lesson will be particularly useful for the string theory discussion in section. \ref{sec:string}. In this case, all the singularities of (\ref{ZNinfty}) are simple poles. If we were to use (\ref{ZNinfty}) to compute $\rho(E)$ with an inverse Laplace transformation, we will pick up a contribution from each of these poles as $\rho(E)\sim e^{\tilde{\beta} E}$ with $\tilde{\beta}$ being the pole. This is another way of seeing the connection between the saddles and the corrections to $\rho(E)$. Of course, the above argument is only good for $E\sim \mathcal{O}(1)$. For $E\sim \mathcal{O}(N^2)$ we can't use (\ref{ZNinfty}) any more. In particular, this simple correspondence between saddle points and singularities do not extend into $E>E_c$ (or more precisely $E+\beta_H\Delta^2 >E_c$). 

All we've discussed can be generalized straightforwardly to theories with fermion excitations. The only difference is that due to $z_{F}(\beta)\neq 0$ in (\ref{BF}), we no longer have $z_{m}(\beta) = z_{1} (m\beta)$.

\subsection{Approximate solutions in the deconfined phase}\label{sec:gapped}

\begin{figure}[t!]
\begin{center}
\includegraphics[scale=0.3]{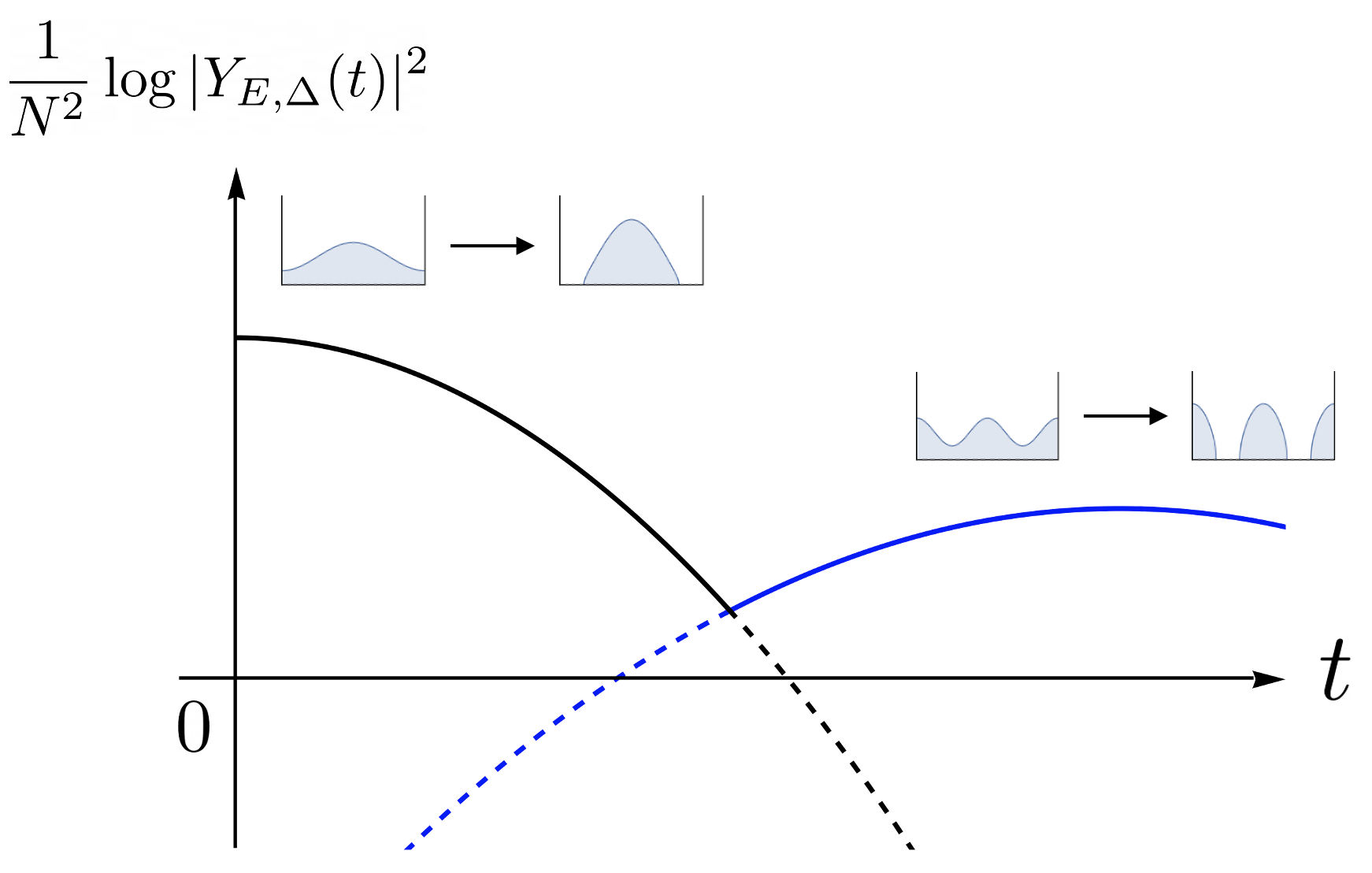}
\caption{The discussion for $E+\beta_H \Delta^2 <E_c$ has a simple continuation into the deconfined regime $E + \beta_H \Delta^2 >E_c$. For example, the saddle point which has a $\cos 2\theta$ variation continues into a two-cut gapped solution. The basic structure of the spectral form factor can remain the same.}
\label{fig:gapphase}
\end{center}
\end{figure}

In this section, we present some preliminary analysis of the story at energy above $E_c$, or above Hagedorn temperature. We leave an analysis of the full model (\ref{Zbeta}) for future, while here we consider a truncated two-term model
\begin{equation}\label{Ztrun}
	Z(\beta) = \int \mathcal{D}U \, \exp\left\{ z_{1}(\beta) \textrm{tr}(U)\textrm{tr}(U^{\dagger }) + \frac{1}{2} z_{2}(\beta) \textrm{tr}(U^2)\textrm{tr}(U^{\dagger2 })\right\},
\end{equation}
and
\begin{equation}\label{Ytrun}
		Y_{E,\Delta}(t) = \int_{\mathcal{C}}  \frac{ \Delta \,d\beta }{\sqrt{2\pi} i}\int   dg_1 dg_2 \,g_1  g_2   \exp\left[ \beta E + \frac{1}{2}\beta^2 \Delta^2 -  \frac{N^2  \,g_1^2}{4z_{1}(\beta +it)} -  \frac{N^2  \,g_2^2}{2z_{2}(\beta +it)} -N^2 F(g_1,g_2) \right].
\end{equation}
This will be a reasonable truncation to do as long as we stay close to $E_c$, or not too far from the Hagedorn transition. To see this, we remind ourselves that the universal behavior of the usual Hagedorn transition is governed by only the first term in (\ref{Ztrun}), while all the higher order terms only give small corrections to the eigenvalue density $\rho(\theta)$. In fact, similar universality extends to the other saddles we discussed in the last section. For each of them, there is a similar ``Hagedorn transition" where in the microcanonical ensemble, the saddle point with $\cos n\theta$ variation transits into a gapped $n$-cut solution. Such a transition will be governed by the $n$-th term in (\ref{ZHS}). Of course, these other transitions are irrelevant for the usual thermodynamics discussion, but they will arise if we look at $Y_{E,\Delta}(t)$ at some fixed $t$ and increase the energy above $E_c$.  This idea is illustrated in fig. \ref{fig:gapphase}.

The conclusion is that the truncated two-term model (\ref{Ytrun}) will suffice in capturing the gapped one-cut early time solution, as well as the gapped two-cut solution which rises and dominates at a later time. We leave the detailed formulas for finding the solutions in appendix. \ref{app:gap}. Since the function $F(g_1,g_2)$ in the gapped phase does not have a very simple form, we have to solve for the saddle points of (\ref{Ytrun}) numerically. As we show in appendix. \ref{app:gap}, we indeed get a curve as advertised in fig. \ref{fig:gapphase}.

\section{Highly excited string gas}
\label{sec:string}

In this section, we discuss the spectral form factor in a different system, a gas of highly excited strings. We will discuss strings in flat space and leave the discussion of strings on curved backgrounds for future study.

Before we embark on the analysis, let us recap on what the main perspective is. In principle, we know what the microstates of a string gas are, and the spectral form factor can be simply computed from it. However, we will consider a different route. We can consider starting from the Horowitz-Polchinski solution \cite{Horowitz:1997jc}, which is a Euclidean classical solution of the winding number $w=\pm 1$ mode on a thermal manifold. One can use the standard Gibbons-Hawking procedure \cite{PhysRevD.15.2752} to compute its entropy, which gives the correct Hagedorn density of states $\rho(E) \sim e^{\beta_H E}$. As we've seen in (\ref{earlyres}), the Hagedorn density of states leads to a rapid decay of $|Y_{E,\Delta}(t)|^2$ and becomes exponentially small at $t\sim \mathcal{O}(1)$. This would be challenging to explain as the spectrum is discrete. Another way to state the problem is that the Horowitz-Polchinski solution only gives a coarse grained density of states, but not the microstates themselves. Note that this is an exact analogue of the problem we have for the Euclidean black hole.

The question is then what we have to add to the Horowitz-Polchinski solution to get the right $|Y_{E,\Delta}(t)|^2$. To answer this question, we go back to the free string limit, where we have better analytic control.\footnote{The Horowitz-Polchinski solution (in the microcanonical ensemble) is specific to four large space-time dimensions. But the free string discussion applies to general cases.} After understanding the important ingredients there, we will then add in weak gravity interaction, and speculate what happens there.

This section can be read independently from section \ref{sec:gauge}. 
However, with no surprise, some lessons we learned from section \ref{sec:gauge} will prove to be quite useful here. Let us review them here. First, in the gauge theory discussion, we saw around eqn (\ref{ZNinfty}) that the complex singularities of the $N\rightarrow \infty$ thermal partition function $Z_{N=\infty}(\beta)$ provide a good grasp on what to anticipate in the spectral form factor $|Y_{E,\Delta}(t)|^2$. In the string theory discussion, this would correspond to the complex singularities of the thermal partition function of a string gas with relatively low energy compared with $1/G_N$, namely that in the free string limit. We denote this thermal partition function of free strings as $Z_{\textrm{free}} (\beta)$. Note that we expect these singularities only because a string gas has a similar Hagedorn density of states as the gauge theory. We are \emph{not} proposing this as a general procedure for other systems as well.

Secondly, in the gauge theory discussion of section \ref{sec:newsaddle}, we saw that the new saddle points in $|Y_{E,\Delta}(t)|^2$ carry nonzero vevs of higher moments of the thermal holonomy $\frac{1}{N}\textrm{tr} (U^n)$. It is well-known that the first moment $\frac{1}{N}\textrm{tr} (U)$, i.e. the Polyakov loop has its analog as the first string winding mode on a thermal manifold, so it is tempting to guess that the new contributions in the string case will have to do with higher winding modes. It turns out that this is in the right direction but not quite. As we will point out, it is true that the new contributions are from other string modes on the thermal manifold, but they do not necessarily carry higher windings.

\subsection{Complex singularities of $Z_{\textrm{free}}(\beta)$}\label{sec:compsing}

The free string thermal partition function $Z_{\textrm{free}}(\beta)$ is convergent when $\textrm{Re}(\beta) > \beta_H$. The first singularity it encounters is $\beta=\beta_H$. However, one can analytically continue $Z_{\textrm{free}}(\beta)$ into other regions of the complex $\beta$ plane and ask whether there are other singularities.
This problem was studied in \cite{Deo:1988jj}. The discussion in this section is mainly a review of their results. 

In this section as well as next sections, we will focus on a highly excited string gas in bosonic string theory in 26 dimensional flat spacetime. The answers in the type II and heterotic string theory will be discussed in Appendix. \ref{app:susy}. 

For concreteness, let's consider the case where all the $25$ spatial directions are compact, with radii $R_i, i=1,2,...,25.$ Physically, we could imagine that some of them $R_1, ...,R_d$ are much larger than string scale, while the rest $R_{d+1}, ..., R_{25}$ are of order $\sim \sqrt{\alpha'}$. Note that there exists a subtlety in the discussion of string thermodynamics, namely that one does not reproduce the answer in noncompact space by simply taking $R_i$ to infinity. The reason is when $R_i$ are finite we have contributions from the winding modes in the spatial directions, which also thermalize at high enough energy even if $R_i$ is large, while they never appear if we started with noncompact dimensions in the first place (see for example \cite{Brandenberger:1988aj} for more discussion on this). Exactly which computation we should use will depend on the real physical set up, but we will not discuss this further here. We simply sum over all the string modes, including the winding modes in the spatial direction. 

 In the limit where string coupling $g_s$ is negligible, the multi-string partition function is given by 
\begin{equation}\label{Zfreebos}
	\log Z_{\textrm{free}} (\beta) = -  \sum_{m}   d(m) \log \left(1 - e^{- \beta m}\right), 
\end{equation}
where $m$ is the energy of the mass of the particle, and $d(m)$ is the degeneracy of mass $m$ in the string spectrum. For bosonic string theory
\begin{equation}\label{levelmatching}
	m^2 = - \frac{4}{\alpha'} + \frac{2}{\alpha'} \left(N_L + N_R\right) + \sum_{i=1}^{25} \left( \frac{n_i^2}{R_i^2} + \frac{w_i^2 R_i^2}{\alpha'^2} \right), \quad \sum_{i=1}^{25} n_i w_i + N_L - N_R = 0
\end{equation}
where $N_L, N_R$ are the left and right oscillator levels and the second equation is the level matching constraint. From now on we will proceed ignoring the usual tachyon in the bosonic spectrum. The lessons can be generalized to supersymmetric cases without tachyons (see Appendix. \ref{app:susy}).

Another useful expression of $	\log Z_{\textrm{free}} (\beta)$ to keep in mind is
\begin{equation}\label{singlestring}
	\log Z_{\textrm{free}} (\beta) = \sum_{r=1}^\infty \frac{z(r\beta)}{r},
\end{equation}
where $z(\beta)$ is the partition function of a single string:
\begin{equation}
	z(\beta) \equiv \sum_{m} d(m) e^{-\beta m}.
\end{equation}
The benefit of (\ref{singlestring}) is that it makes it clear that the singularities in $\log Z_{\textrm{free}}(\beta)$ come from the singularities in the single string partition function $z(\beta)$ and those divided by integer $r\in \mathbb{Z}_+$. This statement has a caveat that there can be cancellations between singularities from different terms in (\ref{singlestring}) \cite{Deo:1988jj}. However, later we will be mostly interested in the leading singularities which come from $z(\beta)$, and for those there is no ambiguity.

As shown by Polchinski \cite{Polchinski:1985zf}, the above expressions (\ref{Zfreebos}), (\ref{singlestring}) can be equivalently derived by considering the torus partition function of a string on a thermal manifold whose Euclidean time has period $\beta$. The same expression can be further cast into the following form \cite{McClain:1986id,PhysRevD.36.1184} (see \cite{Mertens:2015ola} for a nice review)
\begin{equation}\label{fund}
	\log Z_{\textrm{free}} (\beta) = \int_{\mathcal{F}} \frac{d\tau_1 d\tau_2}{ 2 \tau_2} \sum_{\alpha} q^{h_\alpha -1} \bar{q}^{\bar{h}_\alpha-1},
\end{equation}
where $q = e^{2\pi i \tau}, \tau=\tau_1 + i\tau_2$ and the integral is over the fundamental domain $\mathcal{F}$
of $SL(2,\mathbb{Z})$: $\{\tau| |\tau_1| \leq 1/2,\, |\tau|\geq1\}$. In the sum, $\alpha$ labels quantum numbers $\{n,w,N_L,N_R, n_i,w_i\}$ that appear in the conformal weights
\begin{equation}
\begin{aligned}
	& h_\alpha = \frac{\alpha'}{4} \left( \frac{2\pi n }{\beta} + \frac{w \beta}{2\pi \alpha'}\right)^2 + N_L +\frac{\alpha'}{4} \sum_{i=1}^{25}\left( \frac{ n_i }{R_i} + \frac{w_i R_i}{\alpha'}\right)^2, \\
	& \bar{h}_\alpha =  \frac{\alpha'}{4} \left( \frac{2\pi n }{\beta} - \frac{w \beta}{2\pi \alpha'}\right)^2 + N_R+\frac{\alpha'}{4} \sum_{i=1}^{25}\left( \frac{ n_i }{R_i} - \frac{w_i R_i}{\alpha'}\right)^2.
\end{aligned}
\end{equation}
In the formula, $n,w$ are the momentum and winding number in the Euclidean time direction. Note that in (\ref{fund}), we are no longer imposing the level matching condition (\ref{levelmatching}). We rather have $h=\bar{h}$ due to the $\tau_1$ integral. 

It was shown in
\cite{Deo:1988jj} how to analyse the complex singularities of (\ref{fund}) as well as their consequences. Let's focus on the contribution from the term with $\alpha=\{n,w,N_L,N_R, n_i,w_i\}$, 
\begin{equation}\label{Zfree1}
	\log Z_{\textrm{free}} (\beta) \supset \int_{\mathcal{F}} \frac{d\tau_1 d\tau_2}{ 2 \tau_2} \exp\left(-2\pi  B_{\alpha}( \beta)  \tau_2 + 2\pi i C_{\alpha}  \tau_1 \right),
\end{equation}
where
\begin{equation}\label{BC}
\begin{aligned}
	B_{\alpha}( \beta) & = \frac{\alpha'}{2} \left[\left(\frac{2\pi n}{\beta}\right)^2 + \left(\frac{w\beta}{2\pi \alpha'}\right)^2\right]+N_L + N_R -2  + \frac{\alpha'}{2}\sum_{i=1}^{25} \left( \frac{n_i^2}{R_i^2} + \frac{w_i^2 R_i^2}{\alpha'^2} \right),  \\
	C_{\alpha} &   = nw + \sum_{i=1}^{25}n_i w_i  + N_L - N_R.
\end{aligned}
\end{equation}
Only the terms with $C_{\alpha} = 0$ contribute due to the $\tau_1$ integral. For those terms that do contribute, the integral for $\tau_2$ only converges near infinity when $\textrm{Re}(B_{\alpha}(\beta))>0$. This would be true for $\beta$ being a large and positive number. The idea of \cite{Deo:1988jj} is that one could analytically continue ({\ref{Zfree1}}) from the convergent region into the nonconvergent region, leading to an analytic function with a singularity located at places where $B_{\alpha}(\beta) =0, C_{\alpha} = 0$.\footnote{The proper way of thinking about the analytic continuation is to start from the convergent region and continue to the left, going around the singularities one at a time by separating out the divergent piece from the rest (which is convergent there) in the infinite sum (\ref{fund}), and continue the divergent piece as was done in \cite{Deo:1988jj}.  } 

More concretely, (\ref{Zfree1}) leads to a singularity of the form $-\frac{1}{2} \log (B_{\alpha}(\beta))$ and
\begin{equation}\label{sing}
	Z_{\textrm{free}} (\beta) \sim \frac{1}{(\beta - \tilde{\beta})^{\frac{d_\alpha}{2}}},
\end{equation}
where $\tilde{\beta}$ solves $B_{\alpha}(\beta) =0$, and $d_\alpha$ denotes the degeneracies coming from quantum numbers $\alpha$. For example, for the Hagedorn temperature $\beta=\beta_H = 4\pi \sqrt{\alpha'}$, we have $w = \pm 1$ and other quantum numbers being zero, and (\ref{sing}) gives a simple pole  $1/(\beta - \beta_H)$. In fact, $d_\alpha$ will always be even in the current case, since $(n,w) \rightarrow (-n,-w)$ preserves the equations. Therefore all the singularities are poles.\footnote{If some of the spatial directions are non-compact, these singularities are generally branch points.} 

\begin{figure}[t!]
\begin{center}
\includegraphics[scale=0.3]{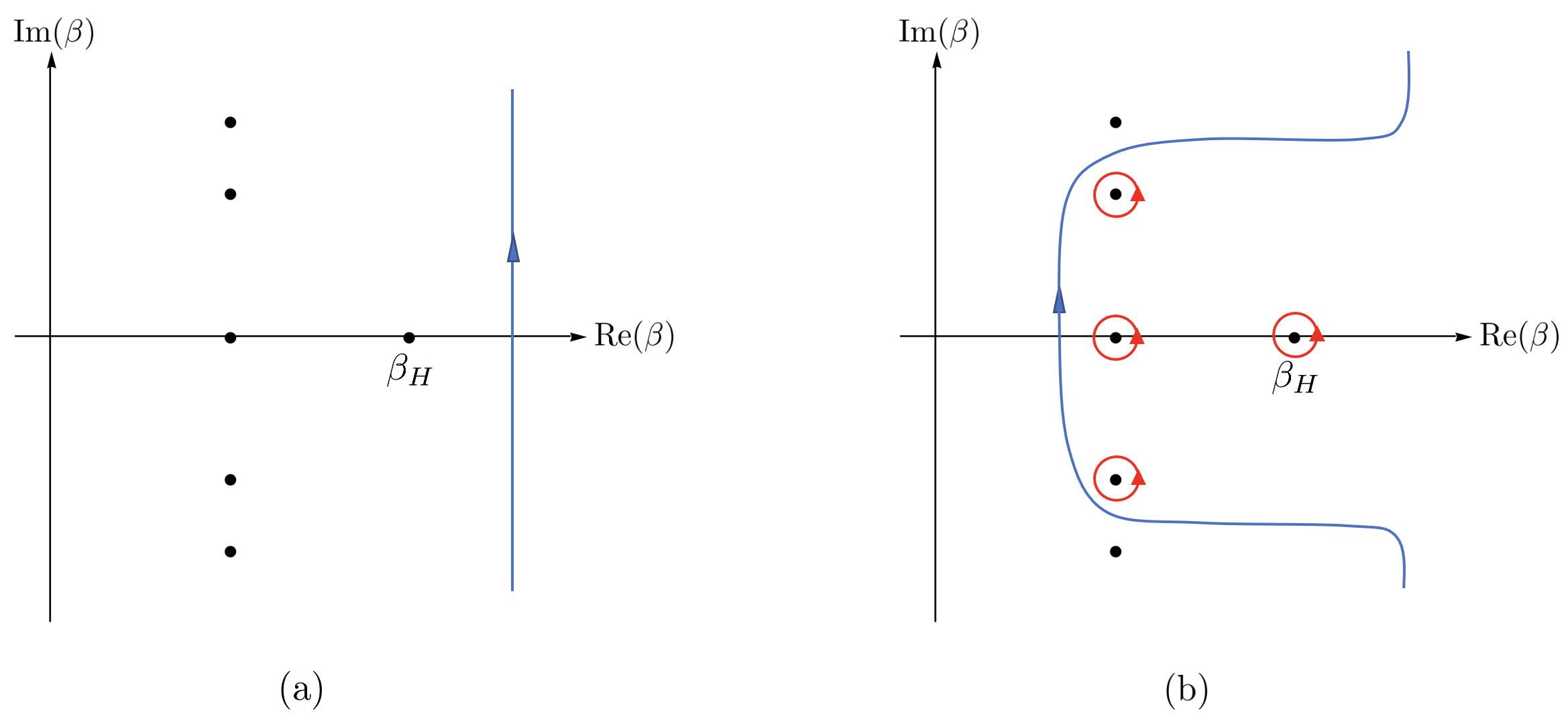}
\caption{(a) The original contour for computing $\rho(E)$. (b) One can deform the contour leftwards, and pick up contributions from the complex poles. For example, the pole at $\beta_H$ simply gives the Hagedorn density of states.}
\label{fig:poles}
\end{center}
\end{figure} 

The reference \cite{Deo:1988jj} then  discussed the consequences of these complex poles in terms of the density of states $\rho(E)$. The idea is that one computes $\rho(E)$ through an inverse Laplace transform of $Z_{\textrm{free}} (\beta)$, by integrating $\beta$ along an imaginary contour whose real part is greater than $\beta_H$. One can then deform the contour leftwards and pick up the contributions from the poles (see fig. \ref{fig:poles} for an illustration). The pole at $\tilde{\beta}$ gives contribution
\begin{equation}\label{stringdensity}
	\rho(E) \sim e^{\tilde{\beta} E},
\end{equation}
with subleading corrections depending on the order of the pole. Since all the other poles have $\textrm{Re}(\beta)<\beta_H$, these are tiny corrections compared with the Hagedorn density of states at high energy. For our later purpose, we also note that the pole at $\tilde{\beta}$ gives a contribution to $Y_{E,\Delta} (t)$ as
\begin{equation}\label{Ystring}
	Y_{E,\Delta} (t) \sim \exp\left[ (\tilde{\beta} -it ) E + \frac{1}{2} (\tilde{\beta} - it)^2 \Delta^2\right],
\end{equation}
whose contribution to the spectral form factor peaks at $t=\textrm{Im}(\tilde{\beta})$.

Before we continue and discuss where the poles reside, we would like to make some comments about the above procedure. In what physical sense can we talk about $Z_{\textrm{free}}(\beta)$ for $\textrm{Re}(\beta) < \beta_H$?  The original definition of the partition function as summing over states does not converge there. The main perspective would be using the analytically continued $Z_{\textrm{free}}(\beta)$ simply as a computational tool, from which we can extract information about the density of states $\rho(E)$, which is completely physical. \cite{Deo:1988jj} used this method to derive various information about $\rho(E)$ which agrees with other methods. In below, we will also check the answer (\ref{stringdensity}) and (\ref{Ystring}) using independent methods.

Another way of thinking about $Z_{\textrm{free}}(\beta)$ is to look at the spacetime effective description \cite{PhysRevD.35.3277,Kogan:1987jd,Atick:1988si}, where we consider an effective theory on  $S^1 \times R^{25}$, with different fields that have masses
\begin{equation}
 m_{\alpha}^2 = \frac{2}{\alpha'} B_{\alpha}(\beta),	
\end{equation}  
where $B_{\alpha} (\beta)$ is given in (\ref{BC}). The singularities of $Z_{\textrm{free}}(\beta)$ correspond exactly to places where some fields become massless. Away from the singularities, the masses are generally complex. At least in the free field approximation, one can define the partition function of the effective theory through analytic continuation from where the masses are real.\footnote{It might be tempting to demand $\textrm{Re}(m_{\alpha}^2) > 0$ as a criterion for the effective theory to be well defined. However, we note that this criterion seems too restrictive. For example, for the first winding mode it requires $\textrm{Re}(\beta^2 - (4\pi\sqrt{\alpha'})^2) >0$, which excludes a large part of the region $\textrm{Re}(\beta)>\beta_H$. On the other hand, we expect the free string partition function to be perfectly well defined there.}

Now, let's discuss where the singularities really are.\footnote{Note that we are artificially ignoring all the singularities with $w=0$, since they come from the usual tachyon in the bosonic string spectrum. I thank Henry Lin for reminding me about them.} For simplicity let's first discuss those with $n_i,w_i =0, i =1,...,25$. They can be added back easily. From $C_{\alpha}=0$ in (\ref{BC}), we get
\begin{equation}\label{nNLNR}
	n = \frac{N_L - N_R}{w},
\end{equation}
This requires $(N_L-N_R)/w\in\mathbb{Z}$. Plug (\ref{nNLNR}) back into $B_{\alpha} (\beta)=0$ in (\ref{BC}) we find
\begin{equation}\label{beta2}
	\frac{\beta^2}{4\pi^2 \alpha'} = \frac{(2-N_L-N_R) \pm 2\sqrt{ (N_L - 1)(N_R - 1) }}{w^2}, \quad n=\frac{N_L - N_R}{w} \in \mathbb{Z}.
\end{equation}

\begin{figure}[t!]
\begin{center}
\includegraphics[scale=0.25]{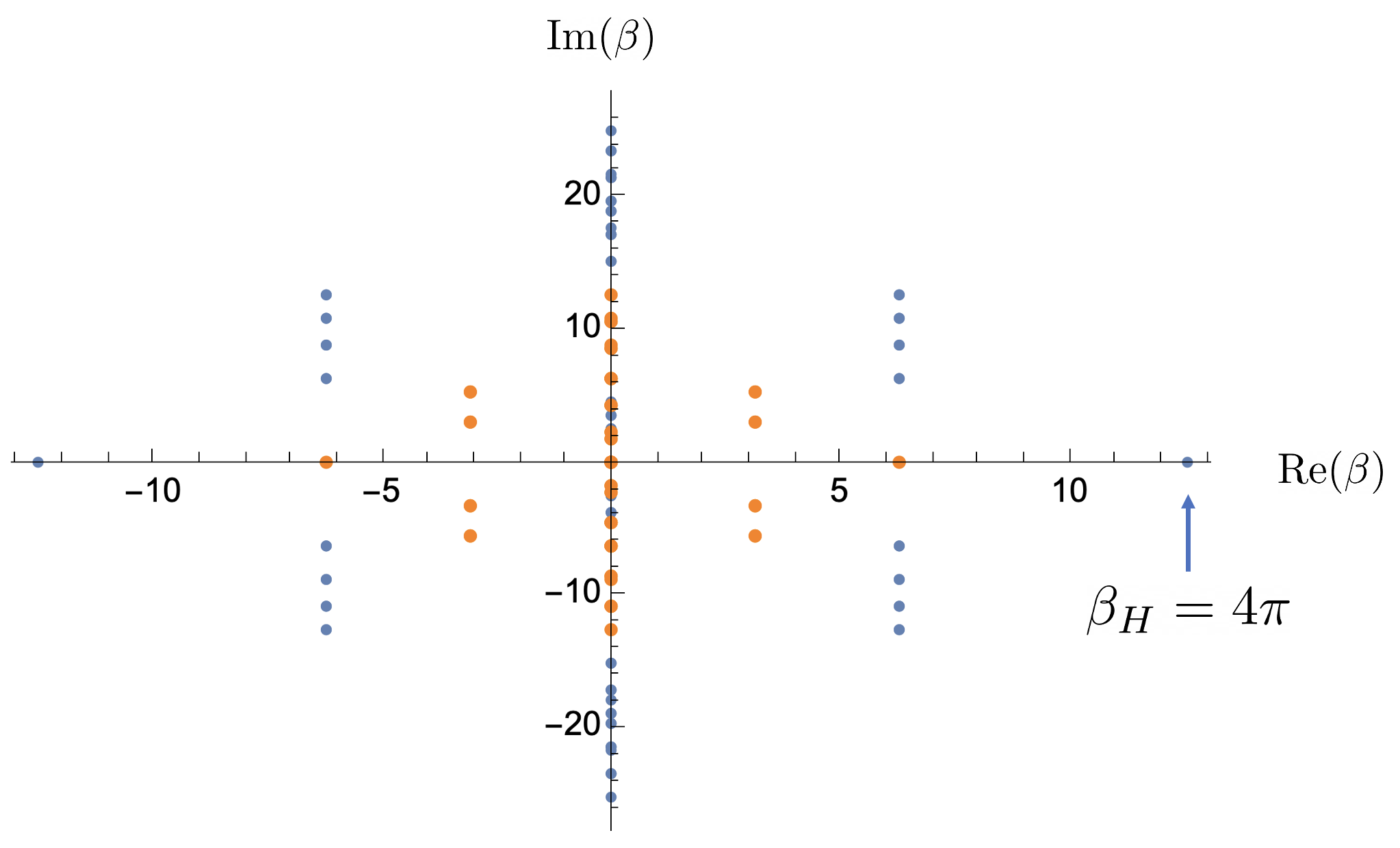}
\caption{We plot the locations of the complex singularities for the bosonic string theory ($\alpha' =1$), with $|w|\leq 2,  N_L\leq 5, N_R\leq 5$. The blue dots have $|w|=1$ and the orange dots have $|w|=2$.}
\label{fig:sings}
\end{center}
\end{figure} 

In fig. \ref{fig:sings}, we plot the singularities with $w= \pm 1,\pm 2, N_L,N_R\leq 5$ on the complex plane. For our purpose of discussing the spectral form factor, we will mainly be interested in the singularities that have greatest real parts, which can be seen from (\ref{Ystring}). It can be seen from (\ref{beta2}) that these singularities (other than $\beta = \beta_H = 4\pi \sqrt{\alpha'}$) are given by
\begin{equation}\label{sin1}
	\frac{\beta}{2\pi \sqrt{\alpha'}} = 1 \pm i \sqrt{N_L - 1}, \quad w = \pm 1,\, n =\pm N_L ,\, N_L > 0,\, N_R = 0,
\end{equation}
or 
\begin{equation}\label{sin2}
	\frac{\beta}{2\pi \sqrt{\alpha'}} = 1 \pm i \sqrt{N_R - 1}, \quad w = \pm 1,\, n =\mp N_R ,\, N_R > 0,\, N_L = 0.
\end{equation}
When $N_L=1$ or $N_R=1$, $\beta = 2\pi \sqrt{\alpha'}$ lies on the real axis and behind $\beta_H$, so they won't be important in the spectral form factor discussion. But we keep them here since this is an often-discussed value that there is an enlarged gauge symmetry (the physical interpretation is only clear when the circle is rather a spatial direction).\footnote{This should also apply to other singularities if we have a spatial circle with complex length.}

Note that at this point, we could already make a prediction for the answer of $|Y_{E,\Delta} (t)|^2$ for a gas of free strings. After the initial decay, it will have many bumps, peaking at $t=2\pi \sqrt{\alpha'}\sqrt{n-1}, n >1$. Note also from (\ref{singlestring}), these singularities already appear in the single string partition function. Therefore we expect a similar behavior for the spectral form factor of a single highly excited string, which we denote by $|y_{E,\Delta}(t)|^2$. We will verify this prediction explicitly in section. \ref{sec:numstring}. 

In the Type II and heterotic string theories, one gets different singularities than (\ref{beta2}), which we discuss in appendix. \ref{app:susy} (the heterotic case was already discussed in the original paper \cite{Deo:1988jj}).

Before we close this section, let's briefly mention the effect of including quantum numbers $n_i,w_i$. Let's consider $n_1,w_1$ as an example.  They can be included by replacing $N_L$ by $N_L +\frac{\alpha'}{4}\left( \frac{ n_1 }{R_1} + \frac{w_1 R_1}{\alpha'}\right)^2$ and $N_R$ by $N_R +\frac{\alpha'}{4}\left( \frac{ n_1 }{R_1} - \frac{w_1 R_1}{\alpha'}\right)^2$ in (\ref{beta2}). They have a chance in competing with the ones in (\ref{sin1}) and (\ref{sin2}) if $R_1 \gg \sqrt{\alpha'}$ and we excite only the momentum modes $n_1$. In such cases, we have
\begin{equation}
		\frac{\beta}{2\pi \sqrt{\alpha'}} = \sqrt{1- \frac{\alpha'n_1^2}{4R_1^2}} \pm i \sqrt{N_L+ \frac{\alpha'n_1^2}{4R_1^2} - 1}, \quad w = \pm 1,\, n =\pm N_L ,\, N_L > 0,\, N_R = 0,
\end{equation}
Still, since the real part of $\beta$ decreases compared to $n_1=0$, we expect them to be subdominant in $|Y_{E,\Delta} (t)|^2$.

\subsection{Spectral form factor for a single highly excited string}\label{sec:numstring}

In this section, we would like to verify some of the claims in the last section using different methods. 

First, we could reproduce (\ref{stringdensity}), namely non-perturbatively small corrections to the density of states, through a more direct computation. In principle we would like to understand it for the multi-string density of states. However, as we mentioned below (\ref{sin2}), the leading singularities (\ref{sin1}) and (\ref{sin2}) appear already in the single-string partition function, so it suffices to look at the density of states for a single string. To further simplify the problem, let's focus on counting the number of oscillator modes, without taking into account the momentum and winding modes in the spatial directions. As we said, we expect their effect to be subdominant. The generating function for the degeneracy of oscillator modes at level $N$ is
\begin{equation}\label{degen}
	\int_{-\frac{1}{2}}^{\frac{1}{2}} d\tau_1 \, \left|\frac{q^{\frac{1}{24}}}{\eta(\tau)}\right|^{48} = 1 + 576 e^{-4\pi \tau_2} + 104976 e^{-8\pi \tau_2} + ...,
\end{equation}
where $\eta(\tau)$ is the Dedekind eta function, $q = e^{2\pi i \tau}, \tau=\tau_1 + i\tau_2$. The $\tau_1$ integral is there to enforce the level matching condition. We follow the usual route to derive the asymptotic behavior for $\rho(N)$ \cite{ramanujan}, i.e. the number of states at level $N$. We have
\begin{equation}
\begin{aligned}
	\rho(N) \propto \int_{\mathcal{C}} d\tau_2\,  e^{4\pi \tau_2 N } \int_{-\frac{1}{2}}^{\frac{1}{2}} d\tau_1 \, \left|\frac{q^{\frac{1}{24}}}{\eta(\tau)}\right|^{48}
\end{aligned}
\end{equation} 
where $\mathcal{C}$ is a contour along the imaginary axis with large real part. We then use the modular transformation of the eta function
\begin{equation}
	\eta(\tau) = \frac{1}{\sqrt{-i\tau}} \eta\left( - \frac{1}{\tau}\right),
\end{equation}
which puts $\rho(N)$ into
\begin{equation}\label{rhoN}
	\rho(N) \propto \int_{\mathcal{C}} d\tau_2\, \int_{-\frac{1}{2}}^{\frac{1}{2}} d\tau_1\, (\tau_1^2 + \tau_2^2)^{12} \exp\left[4\pi \tau_2 (N-1) +\frac{4\pi \tau_2}{\tau_1^2+ \tau_2^2}\right] \frac{1}{\left|\prod_{n=0}^{\infty} \left(1 - \tilde{q}^n \right) \right|^{48}}
\end{equation}
where $\tilde{q} = \exp\left(2\pi i (-1/\tau)\right) = \exp\left(-2\pi \frac{\tau_2 + i \tau_1}{\tau_1^2 + \tau_2^2}\right)$. As usual, when $N\gg 1$, we proceed by looking for saddle points. $|\tilde{q}|$ will be very small, so we can expand the infinite product in (\ref{rhoN}) and start with the leading term, which is simply one. In this case, we get a saddle point at $\tau_1 = 0,\tau_2 = 1/\sqrt{N}$, which leads to
\begin{equation}
	\rho(N) \sim \exp\left(8\pi \sqrt{N}\right).
\end{equation}
This is the usual Hagedorn growth. Now we turn to the subleading terms in the infinite product. Suppose we keep a term of the form $\tilde{q}^{n_l}\bar{\tilde{q}}^{n_r}$, we get an integral
\begin{equation}
	\rho(N) \propto \int_{\mathcal{C}} d\tau_2\, \int_{-\frac{1}{2}}^{\frac{1}{2}} d\tau_1\, (\tau_1^2 + \tau_2^2)^{12} \exp\left[4\pi \tau_2 (N-1) +\frac{4\pi \tau_2}{\tau_1^2+ \tau_2^2} - \frac{2\pi (n_l + n_r) \tau_2}{\tau_1^2 + \tau_2^2} -\frac{2\pi i (n_l - n_r) \tau_1}{\tau_1^2 + \tau_2^2} \right]. 
\end{equation}
The saddle point for $\tau_1$ and $\tau_2$ now locates at\footnote{There are multiple solutions to the saddle point equations, here we focus on the one that gives largest contribution to $\rho(N)$.}
\begin{equation}
\begin{aligned}
	\tau_1 & =  \frac{-i (n_l - n_r)}{2\sqrt{N}\sqrt{(2-n_l - n_r) + 2 \sqrt{(n_l-1)(n_r-1)}} },	\\
	\tau_2 & = \frac{ \sqrt{ (2-n_l -n_r) + 2 \sqrt{(n_l-1)(n_r-1)}   }}{2\sqrt{N}}
\end{aligned}
\end{equation}
and leads to
\begin{equation}\label{rhoNsub}
\rho(N)  \sim \exp\left[4 \pi \sqrt{(2-n_l -n_r) + 2 \sqrt{(n_l-1)(n_r-1)}} \sqrt{N} \right].
\end{equation}
After identifying $n_l,n_r$ with $N_L, N_R$, and $E\sim \frac{2}{\sqrt{\alpha'}} \sqrt{N}$, we get
\begin{equation}
	\rho(E) \sim \exp\left[2 \pi \sqrt{\alpha'}\sqrt{(2-N_L -N_R) + 2 \sqrt{(N_L-1)(N_R-1)}} E\right].
\end{equation}
This reproduces the effect of the singularities in (\ref{beta2}), more specifically, those with winding number $w=\pm 1$ and a plus sign in (\ref{beta2}). In particular, if we set either $n_l$ or $n_r$ to zero, we get the ones in (\ref{sin1}) and (\ref{sin2}) that we are interested in. To see the singularities with $|w|>1$, we have to consider multi-string density of states rather than the single string one.

\begin{figure}[t!]
\begin{center}
\includegraphics[scale=0.28]{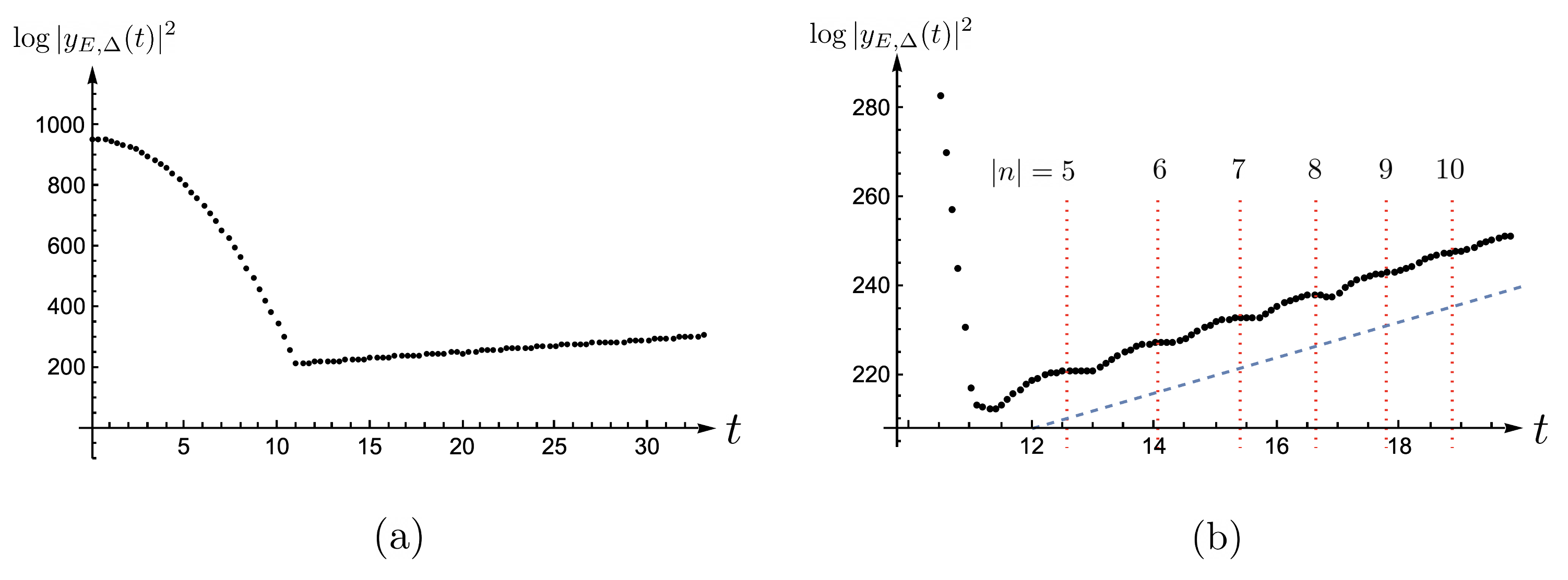}
\caption{(a) We numerically evaluate the spectral form factor for a single string with $E=8/\sqrt{\alpha'}$, $\Delta^2 = 6/\alpha'$ ($\alpha'$ is set to 1 in the figures). This energy is not as low as it might seem. The average energy, rather than being $E$, is actually $E+\beta_H \Delta^2 \approx 83.4/\sqrt{\alpha'}$. Due to numerical reasons, we only considered the contributions from the oscillator modes. (b) We zoom into the latter part of the curve. The red dotted lines correspond to the imaginary parts of $\beta$ in (\ref{sin1}) and (\ref{sin2}), where $n$ labels the momentum number. The blue dashed line has slope $4/\sqrt{\alpha'}$, whose meaning is explained in the main text.}
\label{fig:numstring}
\end{center}
\end{figure}

Needless to say, the above calculation has a huge room for improvement. In particular, if we only have the holomorphic part, it is known due to Rademacher how to improve the asymptotic expansion into a convergent series \cite{Rademacher78} (see a recent discussion \cite{Alday:2019vdr} and references therein). It would be interesting to understand whether one can improve the above calculation along these lines.

Finally, let us present a numerical evaluation of the spectral form factor of a single highly excited string. We denote it by $|y_{E,\Delta}(t)|^2$ to distinguish it from the multi-string answer. The evaluation of the multi-string answer appears numerically demanding to us. We've also further simplified the problem by only focusing on the states with $n_i , w_i = 0$, so all the energy comes from oscillator modes (the degeneracies are simply given by the expansion in (\ref{degen})). This is again mainly due to numerical difficulties. But as we've argued, we should expect to reproduce the effect of the leading singularities (\ref{sin1}), (\ref{sin2}) just from this. 

Indeed, this is what we see from the numerical answer, see fig. \ref{fig:numstring}. In fig. \ref{fig:numstring} (b), we used the red dashed lines to denote the peaks of the spectral form factor predicted from (\ref{sin1}) and (\ref{sin2}). One might be bothered by why the spectral form factor is slowly growing. Since the singularities in (\ref{sin1}) and (\ref{sin2}) have the same real part, naively the peaks should have the same height, according to (\ref{Ystring}). The reason for the rise is actually due to the increasing degeneracies of the winding mode with momentum $n$, which goes like $\sim \exp(4\pi \sqrt{n})$. This should lead to an enhancement factor $8\pi \sqrt{n}$ for $\log |y_{E,\Delta}(t)|^2$ at time $t= 2\pi\sqrt{\alpha'} \sqrt{n-1}$, which predicts a upward linear growth with slope $4/\sqrt{\alpha'}$. We plot a line with this slope in fig. \ref{fig:numstring} (b), offset from the data. We see that it correctly matches the linear growth behavior.

\subsection{Speculations about a family of Horowitz-Polchinski like solution}

What happens if we include interactions, in particular the effect of gravity attraction? In the situation where there are three large spatial directions, this is answered by a classical solution by Horowitz and Polchinski \cite{Horowitz:1997jc}. They considered an effective theory very close to the Hagedorn temperature $\beta_H$, which includes the nearly massless winding number one mode, gravity and dilaton, and found a localized Euclidean solution. See \cite{Chen:2021dsw} for a more detailed review of this solution.  

The Horowitz-Polchinski solution captures the entropy of a self-gravitating string gas in the Lorentzian signature, while it doesn't provide the microstates, or the discreteness of the spectrum. This is much like the situation we have for a Euclidean black hole. Not surprisingly, if we consider the spectral form factor $|Y_{E,\Delta}(t)|^2$ for the Horowitz-Polchinski solution, we get a rapidly decaying answer. We could see this more explicitly as follows. The free energy of the Horowitz-Polchinski solution is given by
\begin{equation}\label{HPaction}
	\beta F = - \log Z =  \frac{\alpha'^{\frac{1}{4}} \xi}{  G_N} (\beta- \beta_H)^{\frac{3}{2}},
\end{equation}
where $\xi$ is some order one number that can be worked out from \cite{Chen:2021dsw}. Now consider the computation of $Y_{E,\Delta}(t)$ 
\begin{equation}
	Y_{E,\Delta}(t) = \int d\beta \,\frac{ \Delta }{\sqrt{2\pi} i}\, e^{\beta E + \frac{1}{2}\beta^2 \Delta^2} Z(\beta+it).
\end{equation}
We will choose both $E,\Delta^2$ to scale as $1/G_N$. One might worry that by taking $\beta$ to $\beta + it$ we are leaving the regime where the Horowitz-Polchinski solution is valid, namely $\beta-\beta_H \ll \sqrt{\alpha'}$. This is a valid concern but we will soon see that the saddle point is such that $\beta +it$ is still close to $\beta_H$. To look for a saddle point, we solve the equation
\begin{equation}\label{saddleHP}
	E + \beta \Delta^2 =  \frac{3}{2} \frac{\alpha'^{\frac{1}{4}} \xi}{  G_N} (\beta + it - \beta_H)^{\frac{1}{2}}.
\end{equation}
Denote the solution at $t=0$ by $\beta_*$, which satisfies $\beta_* - \beta \ll \sqrt{\alpha'}$. We choose both $E$ and $\Delta^2$ to scale as $(\beta_* - \beta_H)^{\frac{1}{2}}$. At time $t$ we make the ansatz $\beta = \beta_* - it + \delta$, and we get
\begin{equation}\label{trange}
	(-it + \delta) \frac{\Delta^2}{(\beta_* - \beta_H)^{\frac{1}{2}} } = \frac{3}{2} \frac{\alpha'^{\frac{1}{4}} \xi}{  G_N} \left[\left(1 + \frac{\delta}{\beta_* -\beta_H} \right)^{\frac{1}{2}}  - 1\right].
\end{equation}
For the solution to be within the approximation by Horowitz and Polchinski, we need $|\delta| \ll \sqrt{\alpha'}$, but we can take it to be larger than $\beta_* - \beta_H$ by some numerical factor. Ignoring the $\delta$ on the left hand side, we get
\begin{equation}
	\delta = (\beta_*-\beta_H) \left[\left(1-i\frac{\gamma t}{\beta_H}\right)^2 -1\right], \quad \gamma \equiv \frac{2G_N\Delta^2 \beta_H}{3 \xi \alpha'^{\frac{1}{4}}(\beta_*-\beta_H)^{\frac{1}{2}}} \sim \mathcal{O}(1).
\end{equation}
So in principle we could trust it until $t$ reaches some large order one number. At that time, the saddle point $\beta + it = \beta_* + \delta$ would have a real part smaller than $\beta_H$ (but still very close). We don't see a clear problem with this since we can in principle analytically continue (\ref{HPaction}) into that regime. If we want to keep the saddle point to the right of $\beta_H$, we can trust it until $t$ reaches order one.

\begin{figure}[t!]
\begin{center}
\includegraphics[scale=0.23]{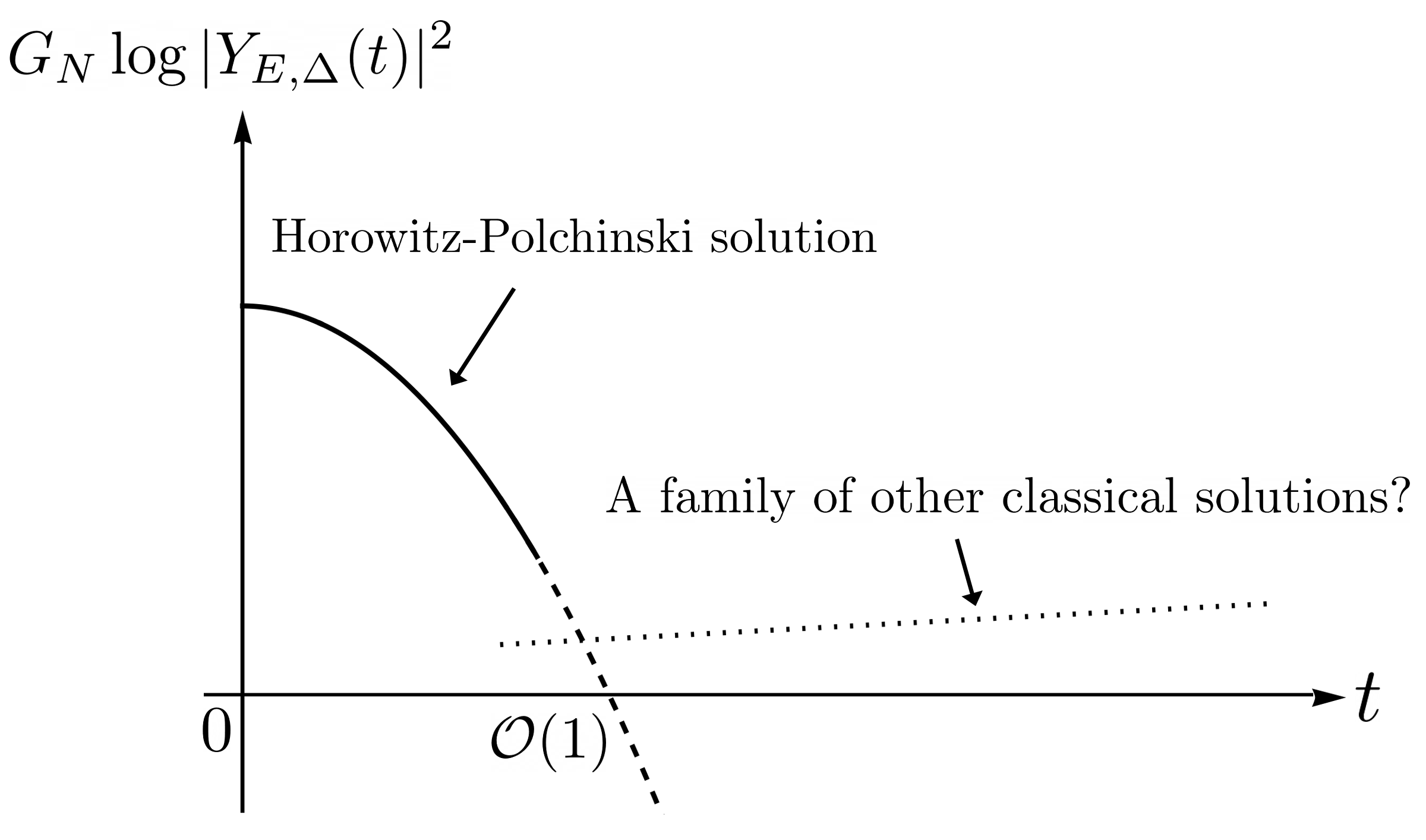}
\caption{The Horowitz-Polchinski solution leads to a decaying contribution for the spectral form factor. The free string discussion suggests that there can be a family of similar classical solutions which are responsible for the rise of the spectral form factor.}
\label{fig:HP}
\end{center}
\end{figure} 

In any case, the saddle point leads to a rapidly decaying spectral form factor:
\begin{equation}
	|Y_{E,\Delta}(t)|^2 \sim \exp\left[ 2 \beta_H E + (\beta_H^2 - t^2) \Delta^2 \right]
\end{equation}
where we only kept the leading terms in the exponential. We sketch it in fig. \ref{fig:HP}.  

In the free string limit, we already see that other winding modes come in, and give a rising behavior to the spectral form factor. What do those contributions in fig. \ref{fig:numstring} become in fig. \ref{fig:HP}? We find the most natural answer to be that there exists a family of Horowitz-Polchinski like solution, one for each of the string modes appearing in the free string case, and they lead to new $\mathcal{O}(\exp(1/G_N))$ contributions to the spectral form factor. 

We find this speculation quite plausible for two reasons. Firstly, physically we don't expect there to be a phase transition between free strings and a string gas with weak interaction. For this reason we expect the qualitative features in the free string answer to remain when we have weak interactions. Secondly, the construction of the Horowitz-Polchinski solution is quite universal. The main important coupling is the coupling between the winding mode and the ``radion" - the mode that characterizes the fluctuation of the Euclidean time circle. Even though the relevant string modes in (\ref{sin1}), (\ref{sin2}) can in general be higher spin fields, the couplings to the radion will be universal when we are close to the complex temperature where they become nearly massless.

Let us take the risk and speculate a bit further on these solutions. To construct these solutions, one would need to consider the theory on a complex spacetime, with length of the Euclidean circle at infinity being close to the point $\tilde{\beta}$ where the string modes in mind become massless. We expect the Euclidean circle to be always finite and close to $\tilde{\beta}$. 
In particular, since the backreaction is small, we expect it to pass the Kontsevich-Segal criterion \cite{Kontsevich:2021dmb,Witten:2021nzp}. 
If we compute the entropy of these solutions using the Gibbons-Hawking procedure, to the leading order in $\beta-\tilde{\beta}$ we should get $S = \tilde{\beta}M$, which reproduces the free string answer (\ref{stringdensity}).

Since $\textrm{Re}(\tilde{\beta}) < \beta_H$, we don't expect the canonical ensemble calculation to be really well defined. We view such construction more as a formal way to construct a solution in the microcanonical ensemble. We've already touched on this question section. \ref{sec:compsing}. In fact, even the Horowitz-Polchinski solution itself contains a negative mode in the canonical ensemble, but it is stable in the microcanonical ensemble.

Of course, there could be other important effects that we haven't taken into account. One might be concerned that the number of string modes that are massless at each singularities grows very fast. This would mean that there are a fair amount of light fields around each singularity, and one might worry about the interaction between these light modes. However, in the original Horowitz-Polchinski solution, the interactions such as the quartic term of the winding modes are not important compared to the gravity attraction. We expect similar argument to still apply here. 

Another issue is that the spacings between the singularities in (\ref{sin1}), (\ref{sin2}) are decreasing with time, as $\delta \tilde{\beta} \sim 1/t$. The Horowitz-Polchinski type solution is only trustworthy when we are close, but not too close to the singularity \cite{Chen:2021dsw}. It breaks down when the quantum fluctuations become large, which happens when the distance from the singularity is of order $g_s^{4/3}$ in four dimensions. This naive argument would suggest that at time of order $t \sim 1/g_s^{4/3}$, we can no longer find a regime where we are close to only one of the singularities while the Horowitz-Polchinski type approximation still applies. But we expect the approximation to hold earlier than this time.

Regardless of the detail, a conceptual point we want to convey is that the winding modes serve as a natural basis in discussing the spectral form factor for a self gravitating gas of string. In a vague sense, winding modes are the ``periodic orbits" for a gravitating string gas.\footnote{I thank Phil Saad for interesting conversations on periodic orbits.} The long time behavior of the spectral form factor might be viewed as coming from an interacting theory of various winding modes. It is an interesting question whether some notion of chaos emerges in this system at late time.

Before we close this section, we would like to point out a puzzling feature about these other winding modes - they carry momentum numbers in the Euclidean time direction. Is there an interpretation of this in the Lorentzian signature?

\section{More discussion on the black holes}\label{sec:bh}

In this section we will return to our starting point in the introduction and discuss some aspects of the spectral form factor of black holes. We will have much less concrete things to say compared to section \ref{sec:gauge} and \ref{sec:string}. In particular, we will only discuss the early time saddle, i.e. the black hole, but not discuss what replaces it.

 The main original point here is to consider the Kontsevich-Segal allowability criterion \cite{Kontsevich:2021dmb,Witten:2021nzp} for complex black holes that appear in the spectral form factors. In section \ref{sec:relation} we also briefly mention some possible connections between black holes and the stories in section \ref{sec:gauge} and \ref{sec:string}.

\subsection{Complex black holes and the allowability criterion}\label{sec:allow}

In \cite{Kontsevich:2021dmb}, Kontsevich-Segal proposed a criterion to determine whether a complex metric is sensible to be included in the gravitational path integral. This is further explained in \cite{Witten:2021nzp} with several interesting examples (see also \cite{Louko:1995jw,Lehners:2021mah}). The idea is to demand that the path integrals of quantum fields, more specifically free $p$-form gauge fields are well defined everywhere on the manifold. With a real basis in the $D$-dimensional tangent space, for a metric that has a diagonal form $g_{ij} = \lambda_i \delta_{ij}$,  the criterion requires that on every point of the manifold,
\begin{equation}\label{allow}
	\sum_{i=1}^D |\textrm{Arg} (\lambda_i)| < \pi.
\end{equation}
In this section we apply this criterion to Schwarzschild black holes, either in flat space or in AdS space.\footnote{The allowability criterion for complex charged/rotating black hole metrics and their applications are being considered by \cite{Gustavo}.} The Euclidean Schwarzschild black hole is real, so it of course passes the criterion (\ref{allow}). However, we will be interested in the black holes that compute $Z(\beta+it)$, and further in computing $Y_{E,\Delta}(t)$.

\begin{figure}[t!]
\begin{center}
\includegraphics[scale=0.28]{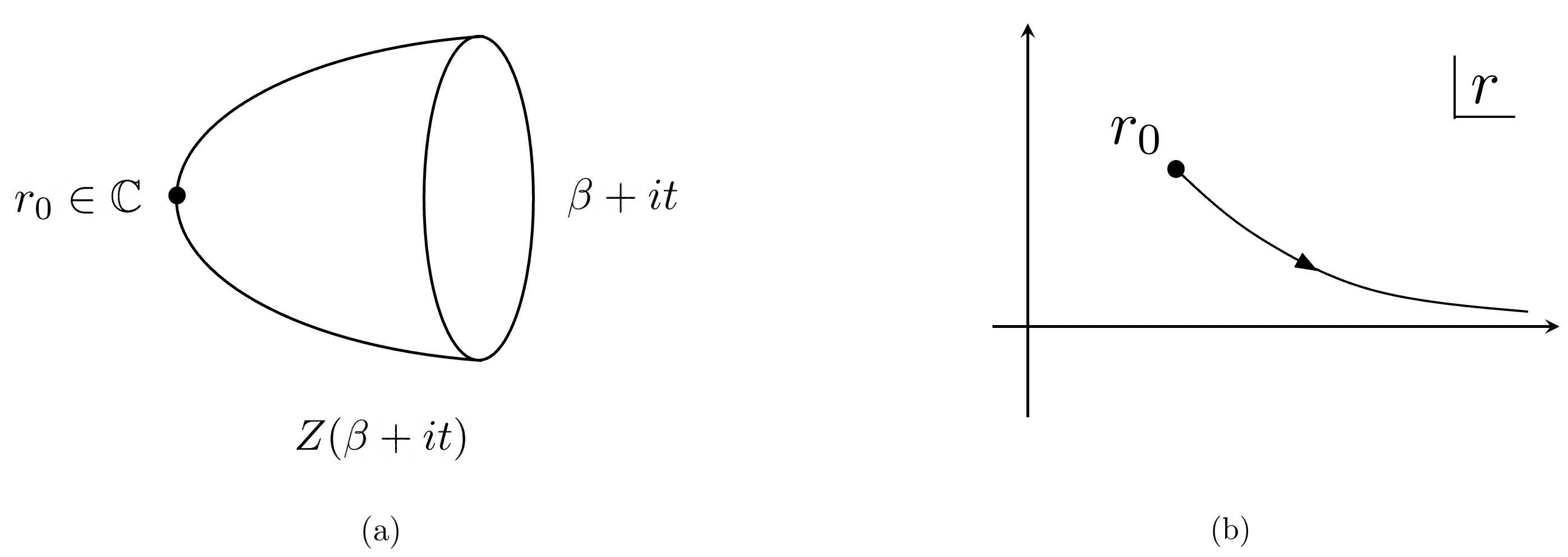}
\caption{(a) The computation of $Z(\beta+it)$ involves a complex black hole. (b) The location of the complex Schwarzschild radius on the complex plane and a contour that connects it to infinity.}
\label{fig:complexBH}
\end{center}
\end{figure}

To compute $Z(\beta + it)$, the boundary length of the black hole is fixed to be $\beta+it$ (or up to a cut off factor in AdS), see fig. \ref{fig:complexBH} (a). Taking flat space  Schwarzschild black hole in $D$ dimensions as an example, the metric for computing $Z(\beta+it)$ is
\begin{equation}\label{Sch}
	ds^2 =  (\beta+it)^2 \left( 1 - \left(\frac{r_0}{r}\right)^{D-3}\right)  d\tau^2 + \frac{dr^2}{1 - \left(\frac{r_0}{r}\right)^{D-3}} + r^2 d\Omega_{D-2}^2,
\end{equation}
where $\tau$ is a \emph{real} parameter, $\tau \sim \tau + 1$. On first sight, this metric seems to always satisfy the criterion (\ref{allow}), since if $r_0$ were real, the only thing that is complex will be from $(\beta+it)^2$, which will have a phase smaller than $\pi$ no matter how large $t$ is. This would have been true for a metric in which the Euclidean circle doesn't shrink, but it is incorrect for the Euclidean black hole, for which the Euclidean circle does shrink to zero and we need to satisfy the smoothness condition at $r=r_0$. The smoothness condition requires
\begin{equation}\label{r0}
	r_0 = \frac{(D-3) (\beta+it)}{4\pi},  
\end{equation}
which means $r_0$ has to be complex as well. (\ref{r0}) is unsurprisingly just the analytic continuation of the usual relation between $r_0$ and $\beta$. 

Since $r_0$ becomes complex, the allowability criterion for (\ref{Sch}) becomes less trivial. To see the constraint it imposes, we can expand (\ref{Sch}) around the horizon
\begin{equation}\label{nearhor}
	ds^2 \approx (\beta+it)^2 \frac{(D-3) (r-r_0)}{r_0} d\tau^2 + \frac{r_0 dr^2}{(D-3) (r-r_0)} + r_0^2 d\Omega_{D-2}^2.
\end{equation}
In the complex $r$ plane, we could approach $r_0$ along a direction such that
\begin{equation}\label{direc}
	\textrm{Arg} \left[(\beta+it)^2   \frac{r-r_0}{r_0}\right] = \textrm{Arg} \left[r_0 (r-r_0)\right] =0,
\end{equation}
where we used (\ref{r0}) in the first equality. Suppose $t>0$, then $\textrm{Arg}(r_0)>0$ and therefore the path of $r$ should approach $r_0$ from the southeast corner, see fig. \ref{fig:complexBH} (b).\footnote{Most of what we say will apply to general black holes, but the direction of the path can vary.} It is easy to see that along such a path near $r_0$, both the first and the second terms in (\ref{nearhor}) will be real, and the only phase will come from the sphere part. The phase from the sphere part is inevitable, and the allowability criterion requires
\begin{equation}\label{argr0}
	2(D-2) |\textrm{Arg}(r_0)|  <\pi. 
\end{equation}
Curiously, (\ref{argr0}) can be written in terms of the ``entropy" of the complex black hole\footnote{This interpretation was suggested to me by Juan Maldacena.} 
\begin{equation}
	\textrm{Re}(S_{\textrm{BH}}) > 0.
\end{equation}
Translating back into (\ref{r0}), we find the criterion for $t$ is
\begin{equation}\label{tcri}
	|t| < \beta \tan \frac{\pi}{2(D-2)}. 
\end{equation} 
Note that this is a very short time, of the thermal scale. This, of course, does not suggest anything wrong with the usual continuation from Euclidean time into real time. In all the usual discussions we always have a Schwinger-Keldysh type contour, where we evolve both forward and backwards in real time. In $Z(\beta+it)$, however, we only evolve in one direction. 

We could repeat the same discussion for AdS Schwarzschild black holes. The metric in $D$ dimensions ($D>3$) is
\begin{equation}
	ds^2 = (\beta+it)^2 \left(\frac{r^2}{\ell^2} + 1- \frac{\mu}{r^{D-3}}\right) d\tau^2 + \frac{dr^2}{\frac{r^2}{\ell^2} + 1- \frac{\mu}{r^{D-3}} } +r^2 d\Omega_{D-2}^2, \quad \mu = r_0^{D-3}\left(\frac{r_0^2}{\ell^2} + 1\right)
\end{equation}
where $\ell$ is the AdS radius. In this case the smoothness condition requires
\begin{equation}\label{adssmooth}
	\beta + it = \frac{4\pi \ell^2 r_0}{(D-1) r_0^2 + (D-3) \ell^2} .
\end{equation}
For a fixed $\beta+it$, (\ref{adssmooth}) has two solutions $r_{\pm}$, which are smoothly connected to the large/small black holes as $t\rightarrow 0$. In either case, we can again look at the criterion coming from the near horizon region. We could again choose to approach the horizon in a direction such that the only phase comes from the sphere part.\footnote{This is always achievable since it is a direct consequence of the smoothness condition, i.e. near the horizon the two dimensional part of the metric can be written as some complex number multiplying $(d\rho^2 +\rho^2 d\theta^2)$.} Then the criterion, in terms of $r_0$, is the same as that in flat space (\ref{argr0}). The criterion in terms of $t$ will be different and is determined by (\ref{adssmooth}), but it is still of the thermal scale. It is also interesting to note that due to the identity
\begin{equation}
	r_+r_- = \frac{(D-3)\ell^2}{D-1},
\end{equation}
the large black hole and the small black hole become unallowable at the same value of $t$.

Above we only used the geometry near the horizon. We still need to be able to find a path to $r=\infty$ that satisfies the criterion everywhere along the path. For Schwarzschild black hole in flat space, one can check that the path that starts at $r_0$, goes straight along the direction (\ref{direc}), and then turns towards infinity after it hits the real axis will be allowed. More generally we have only shown that (\ref{argr0}) is necessary under the Kontsevich-Segal criterion.

For a flat space black hole, we don't have a clear interpretation of (\ref{tcri}). After all, the canonical ensemble in flat space is not well defined. However, (\ref{tcri}) requires an answer in AdS space under the AdS/CFT correspondence. Above the Hawking Page temperature, the answer of $Z(\beta)$ in the CFT is dominated by the large black hole saddle.  If the Kontsevich-Segal criterion is a correct criterion for complex metrics, which has been supported by several examples in \cite{Witten:2021nzp}, we better be able to tell what replaces the black hole in computing $Z(\beta+it)$ of the boundary CFT. 

It turns out that in AdS, the thermal AdS saddle point always dominates over the large black hole \emph{before} the black hole becomes unallowed, so we do not run into a problem.\footnote{The thermal AdS saddle is always allowed since the circle doesn't shrink.} To see this, we use the action difference of an AdS Schwarzschild black hole and thermal AdS \cite{Hawking:1982dh,Witten:1998zw}
\begin{equation}\label{deltaI}
\begin{aligned}
	 I_{\textrm{BH}} - I_{\textrm{thermal AdS}} & = \frac{\textrm{Vol}(S^{D-2}) \left( \ell^2 r_+^{D-2} - r_+^{D}\right)}{4G_N \left((D-1) r_+^2 + (D-3) \ell^2\right)} \\
	& =  \frac{\textrm{Vol}(S^{D-2}) }{4G_N } \left[ - \frac{1}{D-1} r_+^{D-2} + \frac{2D-4}{D-1} \frac{r_+^{D-2} \ell^2}{(D-1) r_+^2 + (D-3) \ell^2}\right].
\end{aligned}
\end{equation}
The reason we put it into the form on the second line is, at the point where the black hole becomes unallowable, $r_+^{D-2}$ is purely imaginary, while the second term in the squre bracket will have positive real part since $r_+^2$ will have the same sign of phase as $r_+^{D-2}$. This means
\begin{equation}
	\textrm{Re} ( I_{\textrm{BH}} - I_{\textrm{thermal AdS}}) >0,
\end{equation}
namely the black hole is subdominant at the point where it becomes unallowable.

This solves the problem in the canonical ensemble quantity $Z(\beta+it)$. Also, we see that even if we started at $\beta < \beta_{\textrm{Hawking-Page}}$, we will soon go into the thermal AdS phase by a first order transition, at $t\sim \mathcal{O}(\beta)$ \cite{Cotler:2016fpe}. This explains the motivation of studying $Y_{E,\Delta}(t)$ instead, a point we already mentioned in the introduction. We've also touched on a similar transition briefly in our free gauge theory discussion. The same type of effect is also studied in \cite{Copetti:2020dil}.

However, this is not the end of the story. If we consider the microcanonical ensemble quantity $Y_{E,\Delta}(t)$, with high enough energy, the thermal AdS saddle will not contribute anymore. In the following section we will see that the black hole saddle is allowable in the microcanonical ensemble even for large $t$.

\subsection{$|Y_{E,\Delta}(t)|^2$ for AdS$_5$ Schwarzschild black holes}

In this section we study the spectral form factor $|Y_{E,\Delta}(t)|^2$ for an AdS black hole. This quantity has been studied for black holes in JT gravity \cite{Saad:2018bqo,Saad:2021uzi}. Here we are just analyzing its behavior in higher dimensions. In particular, we would like to keep track of whether the contributing black hole saddle is allowable or not.

For concreteness, we consider a large Schwarzschild black hole in AdS$_5$ at a very high temperature $\beta\ll \ell$. In this limit, the Euclidean action of the black hole is given by
\begin{equation}
	I_{\textrm{BH}} \approx  - \frac{\pi^2}{8G_N} r_{0}^{3} \approx -\frac{\pi^2}{8G_N} \left( \frac{\pi\ell^2}{\beta+it} \right)^{3}.
\end{equation}
We can check a posteriori that we are always within this approximation. To compute the spectral form factor, we use
\begin{equation}\label{YBH}
	Y_{E,\Delta} (t) = \int d\beta \, \frac{ \Delta }{\sqrt{2\pi} i}\, e^{\beta E + \frac{1}{2} \beta^2 \Delta^2} e^{-I_{\textrm{BH}}},
\end{equation}
and the saddle point equation gives
\begin{equation}\label{sadbh}
	E + \beta \Delta^2 = \frac{3\pi^5 \ell^6}{8G_N} \frac{1}{(\beta+ it)^4}.  
\end{equation}
We denote the solution at $t=0$ by $\beta_*$. Again, both $E$ and $\Delta^2$ are of order $1/G_N$ and $E\sim 1/\beta_*^4, \Delta^2 \sim 1/\beta_*^5$. Note that the amount of $\Delta^2$ is what we will expect for the energy fluctuation in a canonical ensemble at high temperatures.

 We could write the solution at time $t$ as $\beta=\beta_* - it + \delta$, then (\ref{sadbh}) becomes
\begin{equation}\label{tdelta}
	(-it + \delta) \Delta^2 = \frac{3\pi^5 \ell^6}{8G_N} \left[ \frac{1}{(\beta_* + \delta)^4} - \frac{1}{\beta_*^4} \right] .
\end{equation}
Let's consider $t\gg \beta_*$, for which it is consistent to ignore the $\delta$ on the left hand side, and we get
\begin{equation}\label{delta}
	\frac{\delta}{\beta_*} = \frac{1}{\left(1 - i \gamma \frac{t}{\beta_*}\right)^{\frac{1}{4}}} - 1 \approx -1 + e^{i\frac{\pi}{8}} \left(\frac{\beta_*}{\gamma t}\right)^{\frac{1}{4}} ,\quad \gamma \equiv \frac{8G_N \Delta^2 \beta_*^5}{3\pi^5 \ell^6} \sim \mathcal{O}(1).
\end{equation}
Therefore we see that at $t\gg \beta_*$, the boundary length of the black hole becomes $\beta + it = \beta_* + \delta$, which has phase $\textrm{Arg}(\beta_* + \delta) \approx \pi/8$. This means $|\textrm{Arg}(r_0)| \approx \pi/8$, which is within the boundary $\pi/6$ allowed by (\ref{argr0})! 

\begin{figure}[t!]
\begin{center}
\includegraphics[scale=0.24]{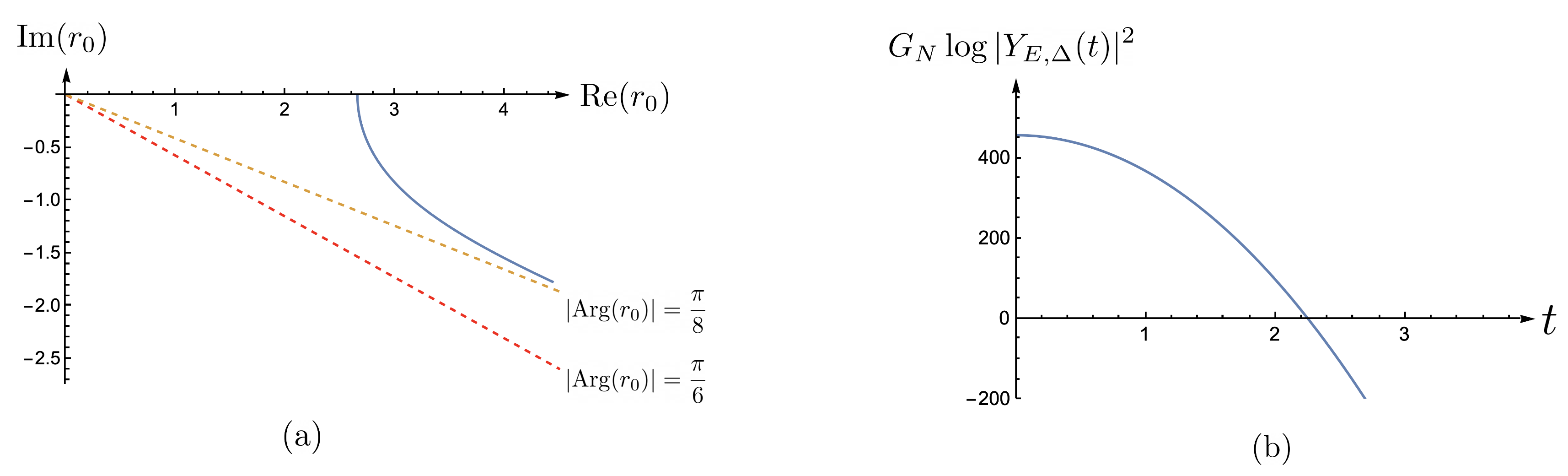}
\caption{In the plots we have $E=\frac{10^2}{G_N},\Delta^2=\frac{10^2}{G_N}$, and the AdS radius $\ell$ is set to one. (a) We show the trajectory of $r_0$ as we increase time. It is always in the allowable region. The boundary of the allowable region is shown by the red dashed line. (b) We show the answer of the spectral form factor given by the semiclassical black hole. It decays to exponentially small at order one time.}
\label{fig:numBH}
\end{center}
\end{figure}

We could go beyond the above approximations by solving the equation (\ref{sadbh}) numerically, with specific parameters. We checked a wide range of parameters and had found that the complex black hole is always allowed. In fig. \ref{fig:numBH} (a) we show an explicit example of the trajectory of $r_0$ on the complex plane.

Finally, we could take the saddle point (\ref{deltaI}) back into (\ref{YBH}), and it leads to a decaying behavior as $\log|Y_{E,\Delta}(t)|^2 \sim \textrm{constant} - t^2 \Delta^2 $. We show an explicit example in fig. \ref{fig:numBH} (b). Note that it soon decays to exponentially small, at an order one time. We've highlighted in the introduction how strange it is from the microscopic theory point of view - $Y_{E,\Delta} (t)$ is a sum of exponentially many order one terms.\footnote{In fact, our original motivation had been to rule out this possibility through the allowability criterion, but this doesn't seem to happen. }

\subsection{Potential ways to connect the free gauge theory/string gas discussion and black holes}\label{sec:relation}

In section. \ref{sec:gauge} and \ref{sec:string}, we've seen how new contributions come in and prevent the spectral form factor from decaying to very small. Of course, both systems are very different from the black hole. However, they are directly related to the black holes in some other limits that were \emph{not} considered in this paper. Let's make some brief comments about them.

\paragraph{Strongly coupled gauge theory and gauge/gravity duality}
Under the gauge/gravity duality \cite{Maldacena:1997re,Witten:1998qj,Gubser:1998bc}, the strong coupling limit of gauge theories are related to black holes. The thermal holonomy $U$ is still a useful quantity at strong couplings. In particular, the Polyakov loop \cite{Polyakov:1978vu,Susskind:1979up}, a localized temporal Wilson loop, is an order parameter for the deconfinement phase transition at finite temperature. The same phenomena can be understood from the bulk as the Hawking-Page transition \cite{Witten:1998zw}. Above the Hawking-Page temperature, the thermal circle shrinks, and a string worldsheet can wrap on the horizon, which signals a nonzero value of the Polyakov loop.

Some authors \cite{Alvarez-Gaume:2005dvb,Alvarez-Gaume:2006fwd} have proposed effective actions of the thermal holonomy $U$ at strong coupling and used them to reproduce the qualitative features of various transitions in the bulk. If we were to use these effective actions, we would expect to encounter some new phases in the spectral form factor. However, there seem to be some more basic questions that we haven't understood. These models usually suggest that the eigenvalue distribution of $U$ in the black hole phase is gapped. However, to our knowledge, there is no clear bulk calculation which suggests this behavior. On the contrary, it seems that in the D0 brane, or BFSS matrix model \cite{deWit:1988wri,Banks:1996vh,Itzhaki:1998dd}, the eigenvalue distribution only has very small fluctuations and does not correspond to a gapped solution \cite{Maldacena:2018vsr}.\footnote{I thank Juan Maldacena for explaining these to me.}

Another way to connect with strong coupling is to consider BPS black holes, which can be studied with matrix models similar to what we considered in section \ref{sec:gauge} (see for example \cite{Cabo-Bizet:2019eaf,Copetti:2020dil,Choi:2021lbk} and references therein). It would be interesting to see whether there is a perspective in studying the microcanonical ensemble quantity (for charges) $Y_{Q,\Delta}$ there as we did.

\paragraph{Black hole/string transition}

We've discussed the spectral form factor for a free gas of string and speculated that under weak gravity attraction, in four dimensions there might be a family of Horowitz-Polchinski like solutions, preventing the spectral form factor from decaying to very small. If we further increase the mass of the string gas, physically it should collapse into a black hole. Therefore, the understanding of the black hole/string transition \cite{Bowick:1985af,Susskind:1993ws,Horowitz:1996nw} will be important to understand how the answer of the string gas extends to the black hole.

However, as was discussed in \cite{Chen:2021dsw}, such a transition cannot be a smooth one as worldsheet CFTs in the type II string theory, based on an index argument (see \cite{Brustein:2021qkj} for a recent discussion in the cigar theory). This is therefore a major obstacle in relating the two sides. In the heterotic theory, there is no known obstruction of a smooth transition, but whether it is actually smooth is still inconclusive \cite{Chen:2021dsw}. 

If it were true that in the heterotic string theory there is indeed a smooth transition, it brings the question what will correspond to the solutions of the other string modes, which we speculated about. Do they correspond to Euclidean black holes with different winding mode profiles near the horizon?\footnote{Even though we are skeptical of this possibility, we should nonetheless mention that the standard no-hair theorems do not apply to winding modes, since their masses become negative near the horizon. To our knowledge no-hair theorems also do not obviously apply to complex black holes.}

\section{Summary and discussion}\label{sec:discussion}

In this paper, we've studied the spectral form factor for free large $N$ gauge theories and free string gases. In both cases, we found new contributions to the spectral form factor after the rapid decay. In the gauge theory, they correspond to new saddle points for the eigenvalue density of the thermal holonomy. For a gas of string, the new contributions come from winding modes which carry some other quantum numbers, in particular momentum in the time direction. In both cases, the new contributions can be identified with non-perturbatively small corrections to the leading continuous density of states.

For the gauge theories, despite the difficulties in connecting these saddle points to the strong coupling limit, which we mentioned in section. \ref{sec:relation}, we do expect that they play an important role when we have weak interactions. It would be interesting to study how the structure gets modified when we have weak interactions, perhaps along the line of the toy models in \cite{Aharony:2003sx}. At weak couplings, a general effective action for computing $Z(\beta)$ can be thought of as an interacting theory of various modes $\frac{1}{N}\textrm{tr}(U^n)$ \cite{Aharony:2003sx}.  An interesting question is how many-body chaos at weak couplings \cite{Stanford:2015owe} can emerge at a longer time from such an interacting theory.

In the case of highly excited strings, we were able to say more about the weakly interacting case due to the Horowitz-Polchinski solution \cite{Horowitz:1997jc}. We've made speculations about a family of classical solutions corresponding to other winding modes that show up in the spectral form factor. It would be interesting to understand this point further.

We note that some specific structures that we saw in the paper, such as complex singularities in the thermal partition function, are closely tied to the Hagedorn density of states in these systems and therefore somewhat special. However, we did provide evidence that the new contributions to the spectral form factor remain even when we go above the Hagedorn temperature in the gauge theories. More generally, any quantum system with a discrete spectrum must have corrections to the continuous density of states and they should manifest themselves in the spectral form factor. 

The surprising finding here is perhaps that the corrections still have good ``geometrical" descriptions and have large (negative) classical actions. In the free gauge theories, they are subleading saddle points but still give answers of order $\mathcal{O}(e^{N^2})$. They also don't have large fluctuations. In the gravity context, analogs of these seem to be geometries that have only one boundaries. For black holes, in general, we don't know of such contributions after the initial decay of the spectral form factor. On the other hand, we do have a double cone wormhole \cite{Saad:2018bqo} which could characterize the spectral form factor under some time average. The classical action of the double cone is zero and therefore seems to give an answer of only order one. If this is what all there is after the initial decay, what is the reason for the absense of $\mathcal{O}(e^{N^2})$ contributions?

In the context of AdS$_3$/CFT$_2$, however, Dyer and Gur-Ari \cite{Dyer:2016pou} have pointed out that a set of $SL(2,\mathbb{Z})$ images of the BTZ black hole contribute to the spectral form factor after the initial decay. Their discussion has many similarities to ours, though we haven't found a sharp way to summarize the common lessons. As a bridge between the two discussions, it might be interesting to generalize our string theory discussion to AdS$_3$ strings \cite{Maldacena:2000kv}.

Finally, a question often asked is the interpretation of the Polyakov loop, or the string winding mode in the Lorentzian signature. We did not answer this in this paper, but we did see that an answer to this question in principle should also involve other moments of the thermal holonomy, or other winding modes.  We note that the spectral form factor can be interpreted as certain transition amplitude \cite{Papadodimas:2015xma,Numasawa:2019gnl,Saad:2019pqd} in the Lorentzian signature. It would be very interesting if there is a physical interpretation of our discussion there.

\paragraph{Acknowledgement}
I'm grateful to Juan Maldacena for discussions and many useful comments. I thank Juan Maldacena and Stephen Shenker for comments on a draft. I thank Raghu Mahajan, Gustavo Turiaci and Edward Witten for discussions about complex black holes. I also thank Ahmed Almheiri, Adam Levine, Henry Lin, Sridip Pal, Phil Saad, Yifan Wang, Zhenbin Yang for interesting discussions.

\appendix

\section{Detail of the equations in the gapped phase}\label{app:gap}

In this appendix, we discuss the equations involved in finding the saddle points in section. \ref{sec:gapped}.  The calculation will be based on the classic reference \cite{Jurkiewicz:1982iz}. Let us copy the action here
\begin{equation}\label{Yapp}
	Y_{E,\Delta}(t) = \int_{\mathcal{C}}  \frac{ \Delta \,d\beta }{\sqrt{2\pi} i}\int   dg_1 dg_2 \,g_1  g_2   \exp\left[ \beta E + \frac{1}{2}\beta^2 \Delta^2 -  \frac{N^2  \,g_1^2}{4z_{1}(\beta +it)} -  \frac{N^2  \,g_2^2}{2z_{2}(\beta +it)} -N^2 F(g_1,g_2) \right].
\end{equation}
We are looking at the truncated model (\ref{Ztrun}), not the full model. As we advertised, the truncated model can give us two relevant solutions (see fig. \ref{fig:gapphase}). The first is the early time solution, corresponding to a gapped one-cut eigenvalue density. The second is a solution which dominates at a later time, which involes two cuts. We discuss them in order.

\begin{figure}[t!]
\begin{center}
\includegraphics[scale=0.24]{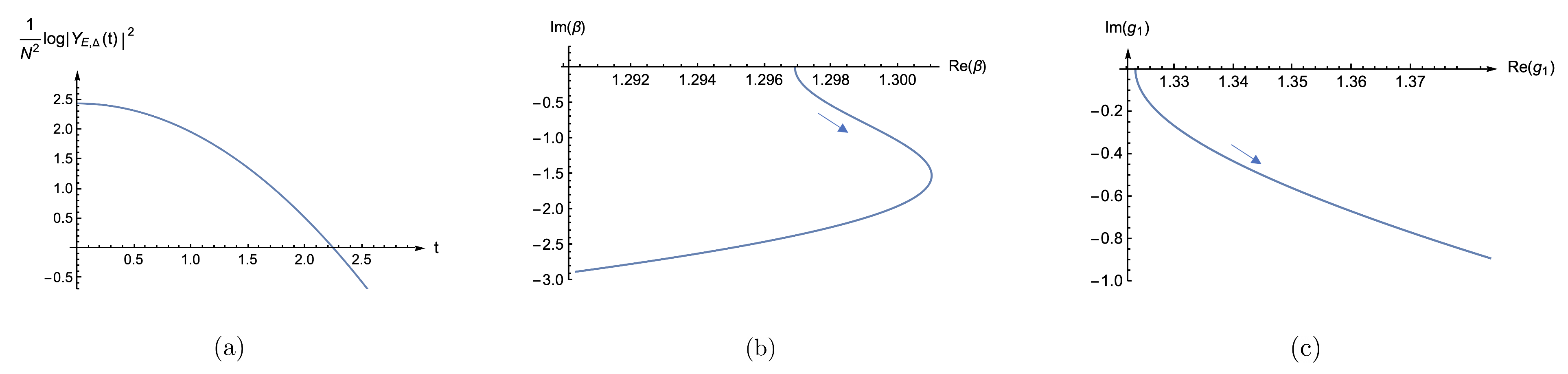}
\caption{The numerical solutions to (\ref{appeqn1}) with $E= 0.6 N^2,\Delta^2 = 0.5N^2$. In (b), (c) we used the arrows to show which directions the curves are running with time. The value of $g_2$ is nonzero in the solution, but it is tiny (of order $\mathcal{O}(10^{-3})$), so we did not show it. }
\label{fig:app1}
\end{center}
\end{figure}

\paragraph{One-cut gapped solution} 
This solution in this truncated model was also considered in \cite{Copetti:2020dil}. The free energy $F(g_1,g_2)$ has the following form (to the leading order in large $N$ expansion)
\begin{equation}
	F(g_1,g_2) = -\frac{1}{4} \left(2 g_2^2 u^4 + u^2 (g_1-4g_2( 
	u - 1) )^2 - 4 (u-1) (-3g_2 u + g_2 + g_1) + 2 \log u \right)
\end{equation}
where
\begin{equation}
	u = \frac{2}{4g_2 + g_1 + \sqrt{(4g_2 + g_1 )^2 - 24 g_2}}.
\end{equation}
The saddle point equations from (\ref{Yapp}) are now
\begin{equation}\label{appeqn1}
\begin{aligned}
 	E + \beta\Delta^2 + \frac{N^2 g_1^2}{4z_1^2} z_1' + \frac{N^2 g_2^2}{2z_2^2} z_2' & =0, \\
    - \frac{ g_1}{2 z_1} - \partial_{g_1} F & =0, \\
    - \frac{g_2}{z_2} - \partial_{g_2} F & =0.
\end{aligned}
\end{equation}
With parameters $E= 0.6 N^2,\Delta^2 = 0.5N^2$ and the model in (\ref{ex1}), we numerically find the solution to the above equation. We show the solution in fig. \ref{fig:app1}. 

\paragraph{Two-cut gapped solution}

\begin{figure}[t!]
\begin{center}
\includegraphics[scale=0.24]{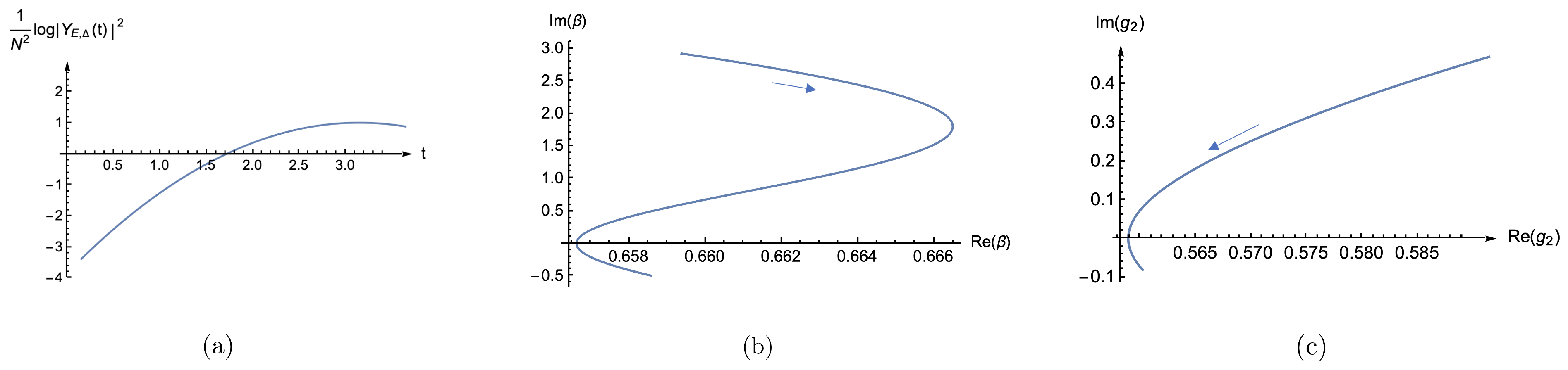}
\caption{The numerical solutions to (\ref{appeqn2}) with $E= 0.6 N^2,\Delta^2 = 0.5N^2$. In (b), (c) we used the arrows to show which directions the curves are running with time. $g_1$ is exactly zero in the solution. }
\label{fig:app2}
\end{center}
\end{figure}

The two-cut gapped solution in the truncated model is actually very simple. The reason is that we can have a saddle point with $g_1$ being exactly zero. This is because if we have a symmetric two-cut solution, as the one we drew in fig. \ref{fig:gapphase}, changing the sign of $g_1$ has no effect since it just shifts $\theta$ by $\pi$. In other words, in this phase we have
\begin{equation}
	\partial_{g_1} F(g_1,g_2)|_{g_1=0} = 0
\end{equation}
and from \cite{Jurkiewicz:1982iz} we know that 
\begin{equation}
	F(0,g_2) = - \left( - \frac{1}{4} \log (2g_2) - \frac{3}{8} + g_2\right).
\end{equation}
If $g_1 =0$, the second equation in (\ref{appeqn1}) is naturally satisfied and we are left with 
\begin{equation}\label{appeqn2}
\begin{aligned}
 	E + \beta\Delta^2  + \frac{N^2 g_2^2}{2z_2^2} z_2' & =0, \\
    - \frac{g_2}{z_2} - \partial_{g_2} F & =0.
\end{aligned}	
\end{equation}
We show the numerical solutions with the same parameters in fig. \ref{fig:app2}. Note that when $t=\pi$, the solution becomes completely real. We note that $\textrm{Re}(\beta)$ is close to $\beta_H/2 \approx 0.658$, which means being close to the Hagedorn transition in this saddle point. 

\section{Complex singularities in superstring theories}\label{app:susy}

In this appendix we discuss the complex singularities of free string thermal partition functions in closed superstring theories at flat space. The heterotic string theory case was already discussed in \cite{Deo:1988jj}. From their answer, one sees that the interesting singularities that will give a large contribution to the spectral form factor lie on the line where $\textrm{Re}(\beta) = 2 \pi \sqrt{\alpha'}$ (in contrast, $\beta_H = \sqrt{2} \pi (1+\sqrt{2})\sqrt{\alpha'}$). 

Here we focus on the Type II string theories. For simplicity, we will focus on the case where all the spatial directions are noncompact. The main complication compared to the bosonic string theory is to incorporate the fact that the spacetime fermions should have anti-periodic boundary conditions along the thermal circle when computing the thermal partition function. The ways to implement this were discussed in a series of papers \cite{Rohm:1983aq,PhysRevD.36.1175,PhysRevD.38.552,Atick:1988si}, and was put into a more genral context in \cite{Kounnas:1989dk}. Here we simply use their the results and do not review the derivation.

We start by writing down the exact expression for the free energy (the same for Type IIA and IIB), as given in (5.20) of \cite{Atick:1988si}:\footnote{We made some modifications compared to the notations in the literature. Our $(w,m)$ maps to $(n,m)$ in \cite{Atick:1988si}, and our $(w,n)$ maps to $(n,m)$ in \cite{PhysRevD.38.552}.}
\begin{equation}\label{freetype2}
\begin{aligned}
	F  \propto & \int_{\mathcal{F}} \frac{d^2 \tau}{\tau_2^6} \left|\frac{1}{\eta(\tau)}\right|^{24} \sum_{w,m} e^{-S_{\beta}(w,m)} \\
	& \times \left[ (\vartheta_2^4 \bar{\vartheta}_2^4 +\vartheta_3^4 \bar{\vartheta}_3^4+\vartheta_4^4 \bar{\vartheta}_4^4 )(0,\tau) +  e^{i\pi (w+m)}  (\vartheta_2^4 \bar{\vartheta}_4^4 +\vartheta_4^4 \bar{\vartheta}_2^4 )(0,\tau) \right. \\ 
	& \quad\quad \left.  - e^{i\pi m} (\vartheta_2^4 \bar{\vartheta}_3^4 +\vartheta_3^4 \bar{\vartheta}_2^4 )(0,\tau) - e^{i\pi w} (\vartheta_3^4 \bar{\vartheta}_4^4 +\vartheta_4^4 \bar{\vartheta}_3^4 )(0,\tau) \right],
\end{aligned}
\end{equation}
where
\begin{equation}
	S_{\beta} (w,m) \equiv \frac{\beta^2}{4\pi \alpha' \tau_2} (m^2 +w^2 |\tau|^2 - 2\tau_1 wm).
\end{equation}
There are several things to explain in (\ref{freetype2}). First, even though $w$ can be identified with the winding number, $m$ is not the momentum. Only after a Poisson resummation on $m$, the new sum will be over the momentum $n$ along the Euclidean time direction. For the terms with an $e^{i\pi m}$ in front, this will result in a sum over $n\in \mathbb{Z}+ \frac{1}{2}$ rather than integers. Second, the various $\vartheta_{\mu}$ functions arise from the one loop partition function of worldsheet fermions, with $\mu$ labeling the spin structure on the worldsheet. $\mu = 1,2,3,4$ corresponds to $\{(+,+), (+,-),(-,-), (-,+)\}$. The first two belong to the R sector, and the last two belong to the NS sector. Finally, the signs of various terms in (\ref{freetype2}) have to do with GSO projections, but we see that they become opposite due to the $e^{i\pi w}$ factors when the winding number is odd. 

We could apply a Poisson resummation on $m$ and bring (\ref{freetype2}) into a form similar to (\ref{fund}). Due to the various signs in (\ref{freetype2}), one might worry about the cancellation between singularities from various terms in (\ref{freetype2}). By using the identity $\vartheta_2^4+\vartheta_4^4-\vartheta_3^4=0$, we can bring (\ref{freetype2}) into a form which is easier to study \cite{Deo:1988jj,PhysRevD.38.552}
\begin{equation}\label{freetype22}
	\begin{aligned}
	F  \propto & \int_{\mathcal{F}} \frac{d^2 \tau}{\tau_2^6} \left|\frac{1}{\eta(\tau)}\right|^{24} \left( \vartheta_2^4 \bar{\vartheta}_2^4 \left(E_0 - E_{\frac{1}{2}} \right) +   \vartheta_4^4 \bar{\vartheta}_4^4 \left(O_0 + O_{\frac{1}{2}} \right) + \vartheta_3^4 \bar{\vartheta}_3^4 \left(O_0 - O_{\frac{1}{2}} \right) \right) ,
\end{aligned}
\end{equation}
where $E_0 , E_{\frac{1}{2}}, O_0 , O_{\frac{1}{2}}$ all have the following form
\begin{equation}
	\sqrt{\frac{\tau_2}{\beta}} \sum_{w,n} \exp\left[ -2\pi i wn \tau_1 - \pi \tau_2 \left( \frac{w^2\beta^2}{4\pi^2\alpha'} + \frac{4\pi^2 n^2 \alpha'}{\beta^2 }\right)\right]
\end{equation}
with the difference being that $E$ ($O$) restricts the winding sum into $2\mathbb{Z}$ ($2\mathbb{Z} + 1$), and subscript $0$ ($1/2$) restricts the momentum sum into $\mathbb{Z}$ ($\mathbb{Z}+\frac{1}{2}$).  

All the terms with momentum being half integers  correspond to the spacetime fermions. From (\ref{freetype22}), it is easy to check that all these terms come with an overall minus sign compared to the spacetime bosons. So instead of leading to singularities in the partition function, they will lead to zeros of the partition function.

\begin{figure}[t!]
\begin{center}
\includegraphics[scale=0.3]{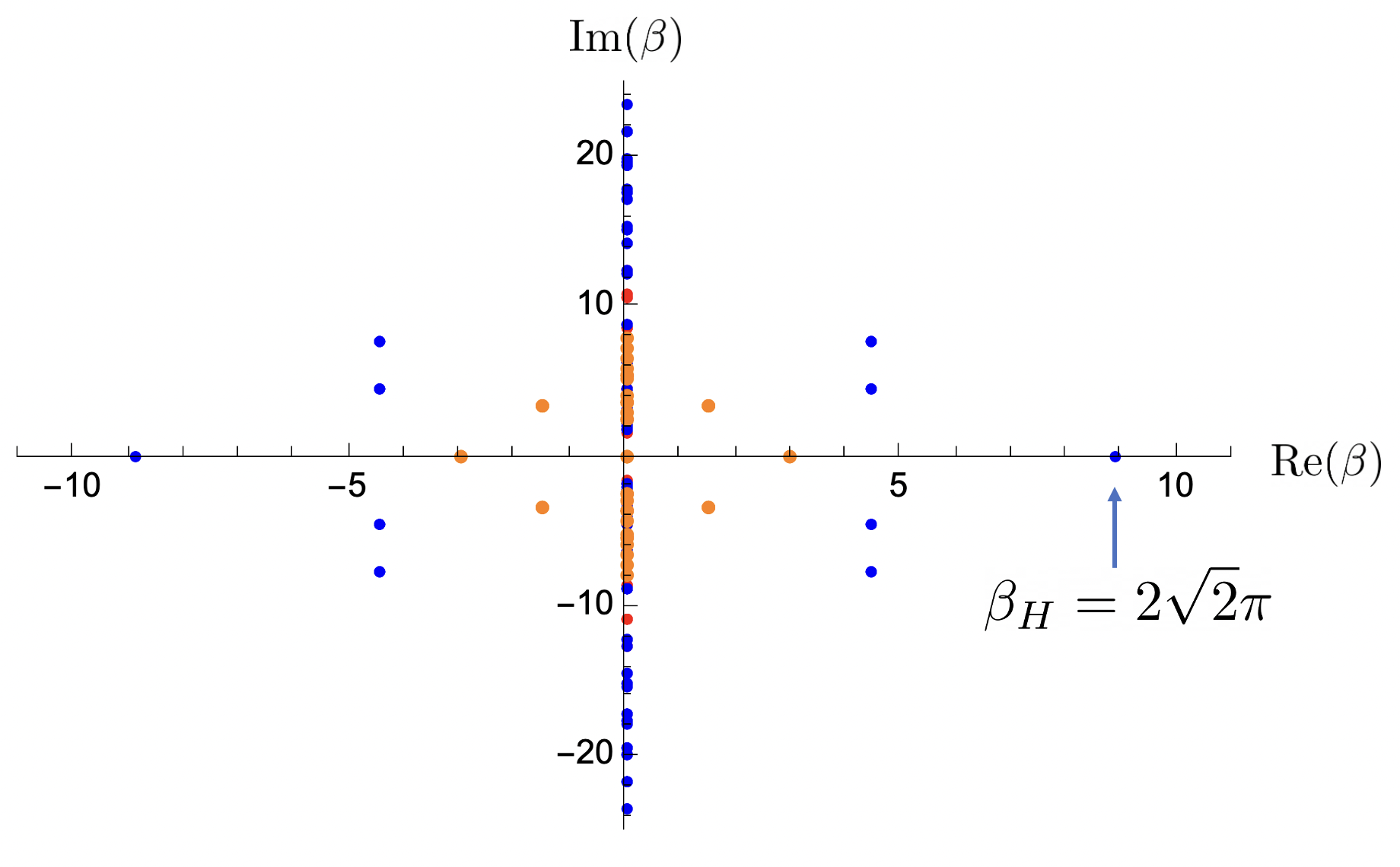}
\caption{We plot the distributions of the complex singularities for Type II string theories ($\alpha' =1$), with $|w|\leq 3, |n| \leq 2, N_L\leq 4, N_R\leq 4$. The blue dots ($w=1$) and the orange dots ($w=3$) correspond to solutions of (\ref{type2eqn2}). The red dots come from (\ref{type2eqn1}). }
\label{fig:type2}
\end{center}
\end{figure}

Focusing on the singularities, which come from the spacetime bosons, we see from (\ref{freetype22}) that possible complex singularities come from two possible sets of equations:
\begin{equation}\label{type2eqn1}
\begin{aligned}
	& \frac{\alpha'}{2} \left[\left(\frac{2\pi n}{\beta}\right)^2 + \left(\frac{w\beta}{2\pi \alpha'}\right)^2\right]+n_l + n_r = 0, \quad n_l - n_r - nw=0,\\
	& \quad w\in 2\mathbb{Z}, \quad n\in \mathbb{Z},\quad n_l,n_r\in \mathbb{Z}_{\geq 0}
\end{aligned}
\end{equation}
corresponding to the first term in (\ref{freetype22}),
 or
 \begin{equation}\label{type2eqn2}
 \begin{aligned}
 & \frac{\alpha'}{2}	 \left[\left(\frac{2\pi n}{\beta}\right)^2 + \left(\frac{w\beta}{2\pi \alpha'}\right)^2\right]+n_l + n_r -1  = 0,\quad n_l - n_r - nw=0, \\
 & \quad  w\in 2\mathbb{Z} + 1, \quad n\in \mathbb{Z}, \quad n_l,n_r\in \frac{1}{2}\mathbb{Z}_{\geq 0}, \quad n_l + n_r \in \mathbb{Z},
\end{aligned}
 \end{equation}
 corresponding to the second and third terms in (\ref{freetype22}). We note that there cannot be a cancellation between (\ref{type2eqn1}) and (\ref{type2eqn2}).  We can lift the restrictions on $n_l + n_r$ in (\ref{type2eqn2})  since otherwise the level matching condition cannot be satisfied. Using these equations, it is easy to work out the singularities. 
 
 In fig. \ref{fig:type2}, we show the singularities with $|w|\leq 3, |n| \leq 2, n_l\leq 4, n_r\leq 4$. The rightermost singularities behind $\beta_H$ corresponds to the solutions of (\ref{type2eqn2}) with $w=1$.

\bibliographystyle{JHEP}
\nocite{}
\bibliography{cite}

\end{document}